\documentclass[prc,twocolumn,aps,tightenlines,showpacs,floatfix]{revtex4}
\usepackage{bm}
\usepackage{color}
\usepackage{graphicx,epsfig}
\usepackage{longtable}

\usepackage{dcolumn}     

\newcommand{\nuc}[2] {$^{#1}$#2}

\def\beq{\begin{equation}}
\def\eeq{\end{equation}}
\def\beqy{\begin{eqnarray}}
\def\eeqy{\end{eqnarray}}

\begin{document}

{

\title{Quantum Monte Carlo calculations of electromagnetic moments and
transitions in $\bm{A\leq9}$ nuclei with meson-exchange currents
derived from chiral effective field theory}

\author{S. Pastore$^1$}
\email{pastore@anl.gov}
\author{\mbox{Steven C. Pieper$^1$}}
\email{spieper@anl.gov}
\author{R. Schiavilla$^{2,3}$}
\email{schiavil@jlab.org}
\author{R. B. Wiringa$^1$}
\email{wiringa@anl.gov}
\affiliation{
$^1$Physics Division, Argonne National Laboratory, Argonne, Illinois 60439\\
$^2$Theory Center, Jefferson Laboratory, Newport News, Virginia 23606\\
$^3$Department of Physics, Old Dominion University, Norfolk, Virginia 23529
}

\date{\today}

\begin{abstract}
Quantum Monte Carlo calculations of electromagnetic moments and
transitions are reported for $A\leq9$ nuclei.
The realistic Argonne $v_{18}$ two-nucleon and Illinois-7
three-nucleon potentials are used to generate the nuclear wave functions.
Contributions of two-body meson-exchange current (MEC) operators are
included for magnetic moments and $M1$ transitions.
The MEC operators have been derived in both a standard nuclear physics 
approach and a chiral effective field theory formulation with pions
and nucleons including up to one-loop corrections.
The two-body MEC contributions provide significant corrections and
lead to very good agreement with experiment.
Their effect is particularly
pronounced in the $A=9$, $T=3/2$ systems, in which they provide up to
$\sim 20\%$ ($\sim 40 \%$) of the total predicted value for the $^9$Li
($^9$C) magnetic moment.
\end{abstract}


\maketitle

}

\section {Introduction}
\label{sec:intro}
Quantum Monte Carlo (QMC) calculations
of electroweak transitions in $A=6,7$ nuclei were reported in Ref.~\cite{PPW07}
and corrections for the magnetic moments (m.m.'s) and $M1$ transitions
from two-body meson-exchange current (MEC) operators were given
in Ref.~\cite{MPPSW08}.
The QMC method is a two-step process, with an initial variational Monte 
Carlo (VMC) calculation to find a good trial function, followed by a 
Green's function Monte Carlo (GFMC) calculation to refine the solution.
When used with the Argonne $v_{18}$ two-nucleon~\cite{WSS95}
and Illinois-2 three-nucleon~\cite{PPWC01} potentials,
the final GFMC results reproduce the ground- and excited-state energies
for $A\le10$ nuclei very well~\cite{PW01,PVW02,PWC04,P05}.

In the present paper, we extend these calculations to $A=8,9$ nuclei using
the improved Illinois-7 three-nucleon potential~\cite{P08b}. The electromagnetic (EM)
current operator includes, in addition to the standard one-body convection and
spin-magnetization terms for individual protons and neutrons, a two-body MEC component.
The latter is constructed within two distinct frameworks, namely the same standard nuclear
physics approach (SNPA) illustrated in Refs.~\cite{Mar05,MPPSW08},
and the chiral effective field theory ($\chi$EFT) formulation of Refs.~\cite{Pastore08,PGSVW09,Piarulli12}.

We report energies, radii, magnetic and quadrupole moments, and
a number of $M1$ and $E2$ transitions.
The MEC contributions can make significant
corrections to the m.m.'s and $M1$ transitions, and we find general
agreement between the two formulations and with experiment.
However the $\chi$EFT formulation provides better agreement
for the calculated m.m.'s, for which both MEC models
are tested. The M1 transitions are calculated only with
the $\chi$EFT MEC operators, showing improved agreement with experiment
in all cases.

A brief review of the QMC calculational method is given in Sec.~\ref{sec:qmc}.
The EM current operator is discussed in
Sec.~\ref{sec:emcnt}.
Results and conclusions are given in Secs.~\ref{sec:res}
and~\ref{sec:concl}.

\section {Quantum Monte Carlo method}
\label{sec:qmc}

We seek accurate solutions of the many-nucleon Schr\"{o}dinger equation
\beqy H \Psi(J^\pi;T,T_z)= E \Psi(J^\pi;T,T_z) \ ,\eeqy
where $\Psi(J^\pi;T,T_z)$ is a nuclear wave function with specific spin-parity 
$J^\pi$, isospin $T$, and charge state $T_z$.
The Hamiltonian used here has the form
\beqy \label{eq:hamiltonian} H = \sum_{i} K_i + {\sum_{i<j}} v_{ij} + \sum_{i<j<k}
V_{ijk} \ ,
\eeqy
where $K_i$ is the non-relativistic kinetic energy and $v_{ij}$ and $V_{ijk}$
are respectively the Argonne $v_{18}$ (AV18) \cite{WSS95} and Illinois-7
(IL7) \cite{P08b} potentials.

The VMC trial function $\Psi_V(J^\pi;T,T_z)$ for a given nucleus is constructed
from products of two- and three-body correlation operators acting on an
antisymmetric single-particle state of the appropriate quantum numbers.
The correlation operators are designed to reflect the influence of the
interactions at short distances, while appropriate boundary conditions
are imposed at long range~\cite{W91,PPCPW97}.
The $\Psi_V(J^\pi;T,T_z)$ has embedded variational parameters
that are adjusted to minimize the expectation value
\begin{equation}
 E_V = \frac{\langle \Psi_V | H | \Psi_V \rangle}
            {\langle \Psi_V   |   \Psi_V \rangle} \geq E_0 \ ,
\label{eq:expect}
\end{equation}
which is evaluated by Metropolis Monte Carlo integration~\cite{MR2T2}.
Here $E_0$ is the exact lowest eigenvalue of $H$ for the specified quantum 
numbers.
A good variational trial function can be constructed with
\begin{equation}
   |\Psi_V\rangle =
      {\cal S} \prod_{i<j}^A
      \left[1 + U_{ij} + \sum_{k\neq i,j}^{A}\tilde{U}^{TNI}_{ijk} \right]
      |\Psi_J\rangle,
\label{eq:psit}
\end{equation}
where the ${\cal S}$ is a symmetrization operator.
The Jastrow wave function $\Psi_J$ is fully antisymmetric and has the
$(J^\pi;T,T_z)$ quantum numbers of the state of interest, while $U_{ij}$
and $\tilde{U}^{TNI}_{ijk}$ are the two- and three-body correlation
operators.
Although we construct the $\Psi_V(J^\pi;T,T_z)$ to be an eigenstate of
the isospin $T$, we allow isobaric analog states with different $T_z$ to have
different wave functions, reflecting primarily the difference in
Coulomb contributions, but also additional charge-symmetry-breaking parts 
of the AV18 interaction.

The GFMC method~\cite{C87,C88} improves on the VMC wave functions by acting
on $\Psi_V$ with the operator $\exp \left[-\left(H - E_0\right)\tau\right]$.
In practice, a simplified version $H^\prime$ of the Hamiltonian $H$
is used in the operator, which includes the isoscalar part of the
kinetic energy, a charge-independent eight-operator projection of AV18 called
AV8$^\prime$, a strength-adjusted version of the three-nucleon potential
IL7$^\prime$ (adjusted so that $\langle H^\prime \rangle \sim \langle H \rangle$),
and an isoscalar Coulomb term that integrates to the total charge of the 
given nucleus.
More detail can be found in Refs.~\cite{PPCPW97,WPCP00}.

The operator is applied in small slices of imaginary
time $\tau$ to produce a propagated wave function:
\begin{eqnarray}
 \Psi(\tau) = e^{-({H^\prime}-E_0)\tau} \Psi_V
          = \left[e^{-({H^\prime}-E_0)\triangle\tau}\right]^{n} \Psi_V \ .
\end{eqnarray}
Obviously $\Psi(\tau=0) = \Psi_V$ and $\Psi(\tau \rightarrow \infty) = \Psi_0$.
The algorithm for propagation produces samples of the wave function
$\Psi(\tau)$ but does not provide gradient information.  
Therefore, quantities of interest are evaluated in terms of a ``mixed'' 
expectation value between $\Psi_V$ and $\Psi(\tau)$: 
\begin{eqnarray}
\langle O(\tau) \rangle_M & = & \frac{\langle \Psi(\tau) | O |\Psi_V
\rangle}{\langle \Psi(\tau) | \Psi_V\rangle},
\label{eq:expectation}
\end{eqnarray}
where the operator $O$ acts on the trial function $\Psi_V$.
The desired expectation values would, of course, have $\Psi(\tau)$ on both
sides; by writing $\Psi(\tau) = \Psi_V + \delta\Psi(\tau)$  and neglecting
terms of order $[\delta\Psi(\tau)]^2$, we obtain the approximate expression
\begin{eqnarray}
\langle O (\tau)\rangle &=&
\frac{\langle\Psi(\tau)| O |\Psi(\tau)\rangle}
{\langle\Psi(\tau)|\Psi(\tau)\rangle}  \nonumber \\
&\approx& \langle O (\tau)\rangle_M
    + [\langle O (\tau)\rangle_M - \langle O \rangle_V] ~,
\label{eq:pc_gfmc}
\end{eqnarray}
where $\langle O \rangle_{\rm V}$ is the variational expectation value.

For the energy, the mixed estimate of Eq.(\ref{eq:expectation}) with
$O = H^\prime$ is itself a strict upper bound to the ground state for
the simpler Hamiltonian, as can be seen by commuting half the imaginary
time operator from the left to right hand side, giving
\begin{eqnarray}
\langle H^\prime(\tau) \rangle_M & = & \frac{\langle \Psi(\tau/2) | 
H^\prime |\Psi(\tau/2) \rangle}{\langle \Psi(\tau/2) | \Psi(\tau/2) \rangle}.
\label{eq:hprime}
\end{eqnarray}
The total energy is then given by this mixed estimate for $H^\prime$ plus
the small difference $\langle (H - H^\prime) \rangle$ evaluated by
Eq.(\ref{eq:pc_gfmc}).

For off-diagonal matrix elements required by transitions the 
generalized mixed estimate is given by the expression
\begin{eqnarray}
&& \frac{\langle\Psi^f(\tau)| O |\Psi^i(\tau)\rangle}{\sqrt{\langle \Psi^f(\tau) | \Psi^f(\tau)\rangle}
\sqrt{\langle \Psi^i(\tau) |\Psi^i(\tau)\rangle}} \nonumber \\
&\approx&
  \langle O(\tau) \rangle_{M_i}
+ \langle O(\tau) \rangle_{M_f}-\langle O \rangle_V \ ,
\label{eq:extrap}
\end{eqnarray}
where
\begin{eqnarray}
\langle O(\tau) \rangle_{M_f} 
& = & \frac{\langle \Psi^f(\tau) | O |\Psi^i_V\rangle}
           {\langle \Psi^f(\tau)|\Psi^f_V\rangle}
      \sqrt{\frac{\langle \Psi^f_V|\Psi^f_V\rangle}
           {\langle \Psi^i_V | \Psi^i_{V}\rangle}} \ , 
\label{eq:mixed_f}  \\
\end{eqnarray}
and $\langle O(\tau) \rangle_{M_i}$ is defined similarly.
For more details see Eqs.~(19-24) and the accompanying discussions in Ref.~\cite{PPW07}.

Sources of systematic error in the GFMC evaluation of operator expectation 
values (other than $H^\prime$) include the use of mixed estimates
and the constrained path algorithm for controlling the Fermion sign problem
in the propagation of $\Psi(\tau)$. These are discussed in Ref.~\cite{WPCP00};
the convergence of the current calculations is addressed at the beginning
of Sec.~\ref{sec:res}.

\section{The electromagnetic current operator}
\label{sec:emcnt}
The nuclear EM current operator is expressed as an expansion in many-body terms.
The current utilized in this work includes up to two-body terms. In what follows,
we use the notation
\begin{eqnarray}
&&{\bf k}_i={\bf p}_i^\prime-{\bf p}_i \ ,\qquad\qquad {\bf K}_i=\left({\bf p}_i^\prime+{\bf p}_i\right)/2 \ ,\\
&&{\bf k}=\left({\bf k}_1-{\bf k}_2\right)/2 \ , \,\qquad {\bf K}= {\bf K}_1+{\bf K}_2 \ ,
\label{eq:ppp}
\end{eqnarray}
where ${\bf p}_i$ (${\bf p}_i^\prime$) is the initial (final) momentum of nucleon $i$, and
${\bf q}={\bf k}_1+{\bf k}_2$ is the momentum associated with the external EM field.

The one-body operator at leading order---or impulse approximation (IA) operator---consists
of the convection and the spin-magnetization currents associated with an individual nucleon.
It is derived from the non-relativistic reduction of the covariant single-nucleon current
by expanding it in powers of ${\bf p}_i/m_N$, where $m_N$ is the nucleon mass and retaining the leading-order term. It reads
\begin{equation}
\label{eq:jlo}
{\bf j}^{\rm IA}=\frac{e}{2\, m_N}
\left[ \,2\, e_{N,1} \, {\bf K}_1
+i\,\mu_{N,1}\, {\bm \sigma}_1\times {\bf q }\,\right] \ ,
 \end{equation}
where
\begin{equation}
e_N = (1+\tau_z)/2 \ , \,\,\,
\kappa_N =  (\kappa_S+ \kappa_V \, \tau_z)/2 \ , \,\,\, \mu_N = e_N+\kappa_N  \ .
\label{eq:ekm}
\end{equation}
Here $\kappa_S= - 0.12$ n.m.\ and $\kappa_V = 3.706$ n.m.\ are the isoscalar (IS) and
isovector (IV) combinations of the anomalous m.m.'s of the proton and neutron,
$e$ is the electric charge, and $\tau_z$ is the Pauli isospin projection equal 
to $+1$ for protons and $-1$ for neutrons.

The calculations of the m.m.'s of the $A\leq 9$ nuclei have been
carried out utilizing two models for the two-body EM current operator,
which are discussed in the next two subsections.

\subsection{$\chi$EFT current operator}
\label{sec:chieft}
\begin{figure}
\includegraphics[height=.25\textheight]{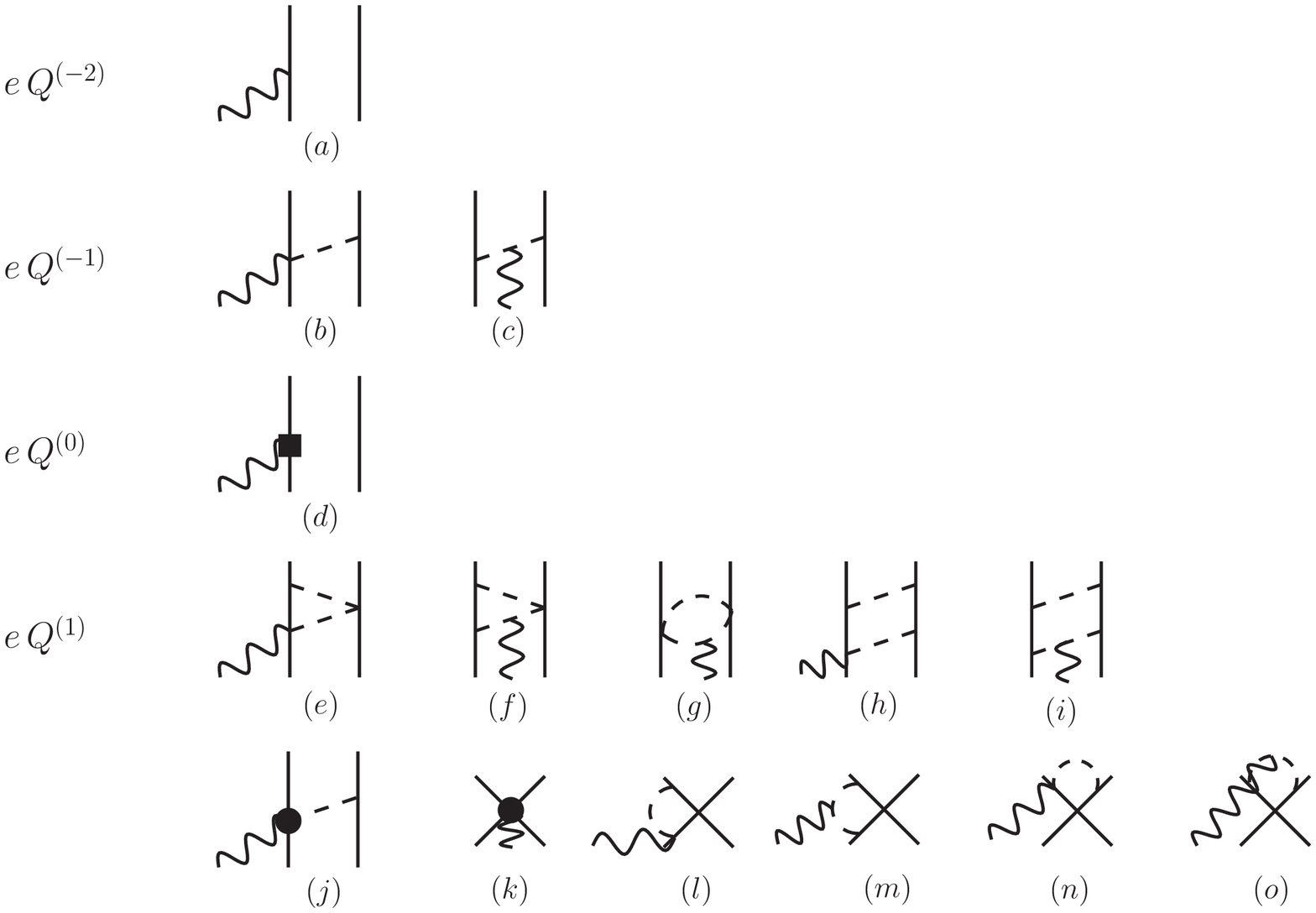}
\caption{Diagrams illustrating one- and two-body $\chi$EFT EM currents
entering at LO ($e\, Q^{-2}$), NLO ($e\, Q^{-1}$), N2LO  ($e\, Q^{\,0}$),
and N3LO ($e\, Q^{\,1}$).  Nucleons, pions,
and photons are denoted by solid, dashed, and wavy lines, respectively.}
\label{fig:f1}
\end{figure}
Two-body EM currents have been derived in recent years within pionful chiral
effective field formulations~\cite{Park96,Pastore08,PGSVW09,Piarulli12,Kolling09,Kolling11}
(for a comparison of the different formalisms we refer to the last four cited references).
Here, we utilize the operators constructed within the formalism developed
in Refs.~\cite{Pastore08,PGSVW09}.
The $\chi$EFT operators are expanded in powers of pions' and nucleons'
momenta, and consist of long- and intermediate-range
components which are described in terms of one- and two-pion exchange contributions,
as well as contact currents encoding short-range mechanisms unresolved
at the given order. These operators involve
a number of low energy constants (LECs) which are fixed to the experimental data.
The operators and fitting procedure have been recently described in Ref.~\cite{Piarulli12}.
We refer to that work for a complete listing of the operators utilized
in the present calculations, and limit ourselves to discussing the various
contributions and to summarizing the fitting strategy adopted to constrain the LECs.
\begin{table}[t]
\caption{Dimensionless values of the isoscalar and isovector LECs entering the
$\chi$EFT current operator at N3LO corresponding to cutoff $\Lambda=600$ MeV.}
\begin{tabular}{c|c|c|c|c|}
$\Lambda$  & $C_{15}^\prime \times \Lambda^4$  & $d_{9}^\prime \times \Lambda^2 $ 
           & $C_{16}^\prime \times \Lambda^4$  & $d_{8}^\prime \times \Lambda^2 $\\
\hline
600  & 5.238 & --0.2033 & --1.025 & 4.980\\
\hline
\end{tabular}
\label{tb:lecs}
\end{table}

The $\chi$EFT EM operators are diagrammatically represented in Fig.~\ref{fig:f1}.
They are expressed as an expansion in the low-momentum scale $Q$.
Referring to Fig.~\ref{fig:f1}, the leading-order (LO) term is counted
as $e\,Q^{-2}$, and corresponds to the IA one-body
operator given in Eq.~(\ref{eq:jlo}). The NLO term (of order $e\, Q^{-1}$)
consists of the seagull and pion-in-flight one-pion-exchange (OPE) currents.
These purely isovector
currents involve two known LECs, the axial coupling constant
$g_A = 1.29$, and the pion decay amplitude $F_\pi=184.6$ MeV.
The value for $g_A$ is determined from the Goldberger-Treiman
relation $g_A=g_{\pi NN}F_\pi/(2\, m_N)$, where the $\pi NN$ coupling constant
is taken to have the value $g^2_{\pi NN}/(4\pi)=13.63 \pm 0.20$~\cite{Stoks93a,Arndt94}.
The N2LO one-body contribution (of order $e\, Q^0$) is a relativistic correction to
the IA operator, and is thus expressed in terms of the nucleons' experimental m.m.'s.

At N3LO ($e\, Q$) we distinguish among four kinds of currents.
The first one (``LOOP'' in the tables), accounts for the one-loop contributions
of diagrams (e)--(i) and (l)--(o). These terms lead to a purely
isovector current which involves the known LECs $g_A$ and $F_\pi$.
For diagram (m) we use the expression given in Ref.~\cite{Piarulli12}
which differs from that given in previous works~\cite{Pastore08,PGSVW09}
by some of the present authors.

Next we account for the contact currents illustrated in panel (k).
We distinguish between minimal (MIN) and non-minimal (NM)
currents. The former is linked to the $\chi$EFT contact potential
at N2LO via current conservation; therefore it involves the same LECs
entering the $N\!N$ potential, and is~\cite{PGSVW09,Piarulli12}
\begin{eqnarray}
{\bf j}^{\rm N3LO}_{\rm MIN}&=&\frac{i\,e}{16}\, \left({\bm \tau}_1\times{\bm \tau}_2\right)_z\,
\Big[ ({\bf k}_1-{\bf k}_2)\nonumber \\
&&\times \left[ C_2+3\, C_4+C_7  +( C_2-C_4-C_7)\, {\bm \sigma}_1\cdot{\bm \sigma}_2
\right]  \nonumber \\
&&+C_7\, [{\bm \sigma}_1\cdot ({\bf k}_1-{\bf k}_2)\, {\bm \sigma}_2+
 {\bm \sigma}_2\cdot ({\bf k}_1-{\bf k}_2)\, {\bm \sigma}_1] \Big] \nonumber\\
&&-\frac{i\, C_5}{4}\, ({\bm \sigma}_1+{\bm \sigma}_2)
\times (e_1\,{\bf k}_1+e_2\, {\bf k}_2)  \ ,
\label{eq:jctm}
\end{eqnarray}
where the low-energy constants $C_1,\dots, C_7$,
have been constrained by fitting $np$ and $pp$ elastic scattering
data and the deuteron binding energy.  We take their values
from the Machleidt and Entem 2011 review paper~\cite{Entem03}.
Unknown EM LECs enter the NM current at N3LO which is given by
\begin{equation}
\label{eq:nmcounter}
{\bf j}^{\rm N3LO}_{\rm NM}= - i\,e \bigg[ C_{15}^\prime\,
{\bm \sigma}_1  + C_{16}^\prime\,
  (\tau_{1,z} - \tau_{2,z})\,{\bm \sigma}_1  \bigg]\times {\bf q}  
+ 1 \rightleftharpoons 2 \ ,
\end{equation}
and the determination of the LECs $C_{15}^\prime$ and $C_{16}^\prime$
is discussed below.

The N3LO OPE current, diagram (j) in Fig.~\ref{fig:f1}, is
given by~\cite{PGSVW09,Piarulli12}
\begin{eqnarray}
{\bf j}^{\rm N3LO}_{\rm OPE} &=& i\, e\, \frac{g_A}{F_\pi^2} \frac{{\bm \sigma}_2 \cdot {\bf k}_2}{\omega_{k_2}^2}\,
\Bigg[ \Big( d_8^\prime \tau_{2,z}+d^\prime_9\, {\bm \tau}_1\cdot{\bm \tau}_2 \Big) {\bf k}_2   \nonumber\\
&&-d^\prime_{21} ({\bm \tau}_1\times{\bm \tau}_2)_z\, {\bm \sigma}_1\times {\bf k}_2  \Bigg] \times {\bf q}
+ 1\rightleftharpoons 2 \ .
\label{eq:cp}
\end{eqnarray}
We fix the LECs multiplying the isovector operators by relating them, in a
resonance saturation picture, to the couplings of the $N$ to $\Delta$ excitation, {\it i.e.},
\begin{equation}
\label{eq:saturation}
\frac{d_8^\prime}{4}\rightarrow \frac{\mu_{\gamma N\Delta} \, h_A}{9\, m_N \,( m_\Delta-m_N)} \ ,
\qquad d^\prime_{21}
= \frac{d_8^\prime}{4} \ ,
\end{equation}
where $m_\Delta$ and $h_A$ are the mass of the $\Delta$ ($m_\Delta-m_N=294$ MeV)
and the $N$ to $\Delta$ axial coupling constant, $\mu_{\gamma N\Delta}=3$ n.m.~is
the transition magnetic moment~\cite{Car86}, and $h_A/F_\pi=f_{\pi N \Delta}/m_\pi$ with
$f^2_{\pi N\Delta}/(4\pi)  \simeq 0.35$ as obtained by equating the first-order expression of the
$\Delta$-decay width with the experimental value.  The current proportional to these
LECs, $d^\prime_8$  and $d_{21}^\prime$, reduces to the
conventional $N$-$\Delta$ current. The isoscalar term
in Eq.~(\ref{eq:cp}) saturates the standard $\rho\pi\gamma$ transition current~\cite{PGSVW09,Piarulli12,Park96}.

The EM operators described above have power-law behavior for large momenta and
need to be regularized before they can be inserted between nuclear wave functions.
The regularization procedure is implemented by means of a cutoff of the form~\cite{Piarulli12}
\begin{equation}
C_\Lambda(k)={\rm exp}(-k^4/\Lambda^4) \ .
\end{equation}

We utilize the $\chi$EFT operators within a hybrid context, in which the
matrix elements are evaluated with wave functions that are
solutions of the realistic Hamiltonian given in Eq.~(\ref{eq:hamiltonian}).
Intrinsic to this approach is a mismatch between the short-range behavior
of the nuclear potential and that of the EM operator.  As a consequence,
the current is not strictly conserved.

The fitting of the unknown LECs entering the EM currents,
namely $C_{15}^\prime$, $C_{16}^\prime$, $d_8^\prime$, and $d_9^\prime$
(with $d^\prime_{21} = {d_8^\prime}/{4}$, as implied by the $\Delta$-saturation mechanism)
was done in Ref.~\cite{Piarulli12}. In that work,
the cut-off $\Lambda$ was varied in the range (500--600) MeV and
the LECs were constrained
to reproduce a set of nuclear EM observables for any given $\Lambda$ in this range.
Three different parametrizations were tested in the $A=2$--$3$ nuclei.
The trinucleon wave functions, required for the evaluation of the matrix elements,
were obtained with the hyperspherical harmonics (HH) expansion discussed in
Refs.~\cite{Kievsky97,Viviani06,Kievsky08} with a nuclear Hamiltonian
consisting of the Argonne $v_{18}$ (AV18) \cite{WSS95} and Urbana IX (UIX)
\cite{PPCW95} potentials. (There is very little difference
between the $A=3$ wave functions for AV18+UIX and AV18+IL7.)
The three models (labeled model I, II, and III) determine the LECs
multiplying isoscalar operators ({\it i.e.} $C_{15}^\prime$ and $d_9^\prime$)
so as to reproduce the experimental deuteron m.m.\ and
the isoscalar combination of the trinucleon m.m.'s.
Models II and III fix the isovector LEC $d_8^\prime$ by $\Delta$-saturation
as indicated in Eq.~(\ref{eq:saturation}), while $C_{16}^\prime$ is constrained
so as to reproduce either the $np$ radiative capture cross section at thermal neutron
energies in model II, or the isovector combination of the trinucleon magnetic
moments in model III.
In model I, $d_8^\prime$ is left as a free parameter and is constrained, along with
$C_{16}^\prime$, so as to reproduce both the $np$ radiative capture cross section
and the isovector combination of the trinucleon m.m.'s.
As already observed in Ref.~\cite{Piarulli12}, model I leads to unnaturally
large values for both isovector LECs, severely spoiling the convergence pattern
of the chiral expansion.  We have, nevertheless, tested all three models
(with cutoffs of both 500 and 600 MeV) in VMC
m.m.\ calculations for $A=3$--$8$ nuclei, and verified that this 
pathology---{\it i.e.}, the lack
of convergence---persists, and indeed gets worse, in larger systems.  We have
therefore disregarded model I, and adopted model III with cut-off $\Lambda=600$ MeV
in the present study.  The parameters entering this
model, obtained from the calculations performed in Ref.~\cite{Piarulli12},
are listed in Table~\ref{tb:lecs}.  Model III (with $\Lambda=600$ MeV), 
when tested in VMC calculations,
produced the best results for the m.m.'s.

\subsection{SNPA current operator}
\label{sec:snpa}

The two-body currents in the SNPA formalism
have been described in detail most recently in Ref.~\cite{Mar05}.
These currents are separated into model-independent (MI) and
model-dependent (MD) terms.  The former (MI) are derived from the $N\!N$
potential (the AV18 in present case), and their longitudinal components
satisfy, by construction, current conservation with it.
They contain no free parameters, and their short-range behavior is
consistent with that of the potential.  The dominant terms, isovector
in character, originate from the static part of the potential, which is assumed
to be due to exchanges of effective pseudoscalar (PS or ``$\pi$-like'') and vector
(V or ``$\rho$-like'') mesons.  The associated currents are then constructed by
using the PS and V propagators, projected out of the static potential~\cite{Mar05}.
Additional (short-range) currents follow by minimal substitution in the
momentum-dependent part of the potential.  They have both isoscalar and
isovector terms, and lead to contributions which are typically much smaller
(in magnitude) than those generated by the PS and V currents.  At large
inter-nucleon separations, where the $N\!N$ potential is driven by the OPE
mechanism, the MI current coincides with the standard seagull and pion-in-flight
OPE currents diagrammatically illustrated in panels (b) and (c), respectively, of
Fig.~\ref{fig:f1}.

The MD currents are purely transverse, and unconstrained by current
conservation.  The dominant term is associated with excitation of
intermediate $\Delta$ isobars, which are treated non-perturbatively
with the transition-correlation-operator method developed in Ref.~\cite{Schiavilla92}.
These $\Delta$ currents are discussed in considerable detail in Ref.~\cite{MPPSW08}.
Additional (and numerically small) MD currents arise from the isoscalar $\rho\pi\gamma$
and isovector $\omega\pi\gamma$ transition mechanisms.  The values for the coupling constants
entering them are also listed in Ref.~\cite{MPPSW08}.

%
\section {Results}
\label{sec:res}
\begin{figure}[hbt]
\includegraphics[width=3.30in]{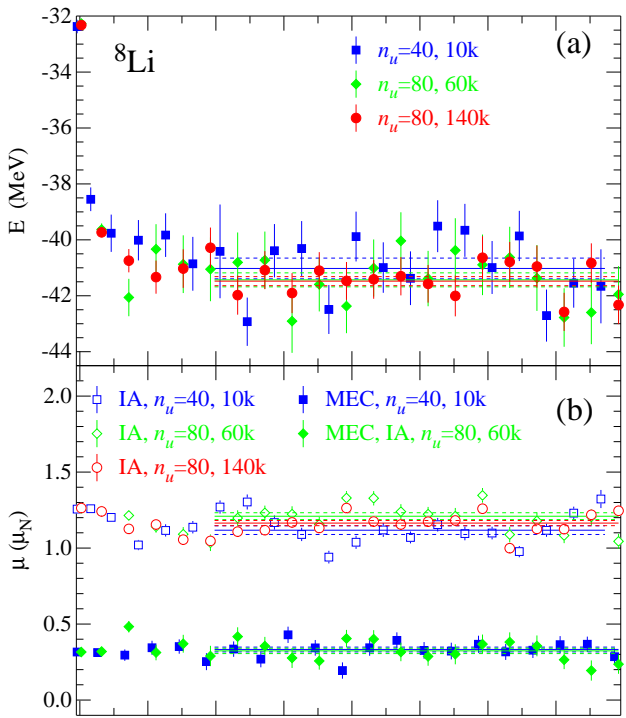}
\caption{(Color online) Propagation of the energy (a) and m.m.\ (b) 
as a function of imaginary time $\tau$ for the ground state of $^8$Li.}
\label{fig:li8_vs_tau}
\end{figure}

The IA m.m.\ for the 8- and 9-body nuclei have significantly higher
Monte Carlo statistical errors than most quantities that we have
computed with GFMC.  Therefore we present two examples of the GFMC
propagation as a function of the imaginary time $\tau$.
Fig.~\ref{fig:li8_vs_tau} shows the propagation of a typical case,
$^8$Li.  Three propagations are shown, one in which the constrained
propagation~\cite{WPCP00} is relaxed with 40 unconstrained steps
($n_u=40$) and two with $n_u=80$.  The $n_u=40$ case was made with
10,000 walkers while the $n_u=80$ cases are averages of calculations with
20,000 and 40,000 (60,000 total) and 20,000, 40,000 and 80,000 walkers (140,000 total),
respectively.  The energy is
shown in panel (a); it is similar to results shown in
Ref.~\cite{WPCP00}.  As can be seen, there is a rapid drop from the
initial VMC value at $\tau = 0$ that reaches a stable result before
0.1 MeV$^{-1}$.  The results for all quantities presented in this
article are averages over $\tau$ from 0.2 to 0.8 MeV$^{-1}$, as
indicated by the solid lines, with statistical errors denoted by the
dashed lines.  The propagations of point proton and neutron radii are
similar, except that the starting VMC values are within 5\% of the final
results.  As has been shown in Ref.~\cite{WPCP00}, these quantities are
all converged by $n_u=40$, often by $n_u=20$.  The quadrupole moments
are much more difficult to evaluate, because they have long-lived
oscillations in the propagation time $\tau$.

The IA and $\chi$EFT MEC m.m.\ are shown in panel (b).  The
statistical fluctuations of the IA term are much worse than those of the
MEC term when the same number of configurations are used.  Also there
may be a small systematic change in the IA term going from $n_u=40$ to
$n_u=80$; the average values are 1.120(27) and 1.164(17), respectively,
giving a difference of 0.044(32).  The MEC does not have this
sensitivity; the two calculations are in excellent agreement.  However
because of the rapid growth of statistical error with increasing $n_u$,
the $n_u=80$ calculation needs seven times as many walkers to achieve
the same statistical error.

\begin{figure}[bth]
\includegraphics[width=3.30in]{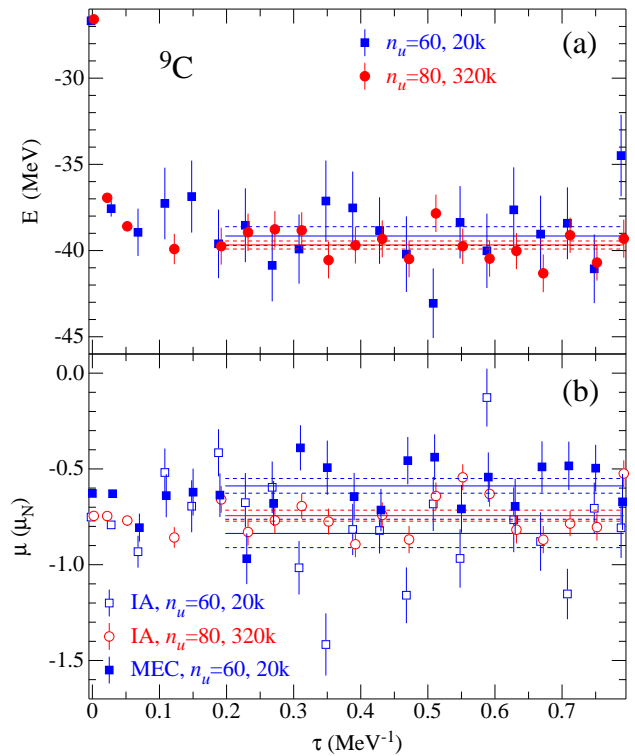}
\caption{(Color online) Propagation of the energy (a) and m.m.\ (b) 
as a function of imaginary time $\tau$ for the ground state of $^9$C.}
\label{fig:c9_vs_tau}
\end{figure}

The Monte Carlo statistical errors in our computed m.m.'s are 
most severe for $^9$C, as is shown in Fig.~\ref{fig:c9_vs_tau}.
Two propagations are shown, one in with $n_u=60$ and one with  $n_u=80$.
The first was made using 20,000 walkers while the latter is the average of
two calculations, each with 160,000 walkers.  Again the energy, shown in
the panel (a), is well converged for both cases.  As before for $^8$Li, the $\chi$EFT MEC m.m., panel (b),
has smaller fluctuations for a given number of walkers than
the IA m.m.  In this case the $\chi$EFT MEC is a large (80\%) addition to
the IA m.m.  

In both examples the statistical fluctuations in the IA term are much
larger than in the MEC term.  However the evaluation of the MEC requires
much more computational effort per walker than does the propagation
and IA term.  Therefore for most of the calculations, we propagate 
a large number of walkers using $n_u=60$ or 80 to obtain the IA m.m.\ term
(and also the other reported quantities).
The MEC is obtained with comparable statistical error using fewer walkers
and the two numbers and their errors combined to get the total m.m.
The propagations are averaged over $\tau = 0.2$ to 0.8~MeV$^{-1}$.

The large statistical fluctuations (and possible $n_u$ sensitivity) are
coming from the IV combination of the spin term of the IA
m.m., Eq.~(\ref{eq:jlo}). The IV convection term and both IS terms have
much smaller fluctuations.  Thus if isospin symmetry is assumed for the
wave functions of isobaric analogs, we can make precise statements about
the IS m.m.  However if we do not want to assume such isospin symmetry,
as in the $^9$C--$^9$Li case below, then we have to make separate
calculations for each nucleus and the large errors in the IV parts make
the extraction of an IS m.m.\ with small statistical error impossible.

\begin{table*}[bth]
\caption{GFMC results for $A\leq9$ nuclear states studied in this work,
compared to experimental values~\cite{Audi03,Purcell10,Tilley02,Tilley04,
Borremans05,Mohr12,Amroun94,Shiner95,Nortershauser11,Nortershauser09}. 
Numbers in parentheses are statistical errors for the GFMC calculations
or experimental errors; errors of less than one in the last decimal place
are not shown.}
\label{tb:energies}
\begin{ruledtabular}
\footnotesize
 \begin{tabular}
{lD{.}{.}{3.3}D{.}{.}{3.1}D{.}{.}{3.9}D{.}{.}{3.3}D{.}{.}{3.3}D{.}{.}{3.3}D{.}{.}{3.1}D{.}{.}{3.3}}
   $^AZ(J^\pi,T)$                &
   \multicolumn{2}{c}{$E$ [MeV]} &
   \multicolumn{2}{c}{$r_p$[$r_n$] [fm]} &
   \multicolumn{2}{c}{$\mu$(IA) [n.m.]} &
   \multicolumn{2}{c}{$Q$ [fm$^2$]}
   \\ \cline{2-3} \cline{4-5} \cline{6-7} \cline{8-9}
                            &
   \multicolumn{1}{c}{GFMC} &
   \multicolumn{1}{c}{Expt.}&
   \multicolumn{1}{c}{GFMC} &
   \multicolumn{1}{c}{Expt.}&
   \multicolumn{1}{c}{GFMC} &
   \multicolumn{1}{c}{Expt.}&
   \multicolumn{1}{c}{GFMC} &
   \multicolumn{1}{c}{Expt.}
   \\ \hline
   \nuc{2}{H}$(1^{+},0)$                    &  -2.225   &  -2.2246& 1.968     & 1.976(3) & 0.847 & 0.8574 & 0.270   &  0.286
   \\
   \nuc{3}{H}$(\frac{1}{2}^+,\frac{1}{2})$  &  -8.50(1) &  -8.482& 1.58 [1.76]& 1.58(10)& 2.556 & 2.979
   \\
   \nuc{3}{H}$(\frac{1}{2}^+,\frac{1}{2})$* &  -8.46(1) &        & 1.60 [1.80]&         & 2.550 &
   \\
   \nuc{3}{He}$(\frac{1}{2}^+,\frac{1}{2})$ &  -7.73(1) &  -7.718& 1.80 [1.60]& 1.76(1) &-1.743 &-2.127
   \\
   \nuc{3}{He}$(\frac{1}{2}^+,\frac{1}{2})$*&  -7.75(1) &        & 1.76 [1.58]&         &-1.750 &
   \\
   \nuc{6}{Li}$(1^+,0)$                     & -31.82(3) & -31.99 & 2.39       & 2.45(4) & 0.817 & 0.822  &-0.20(6) &-0.082(2)
   \\
   \nuc{6}{Li}$(0^+,1)$                     & -28.44(4) & -28.43 &
   \\
   \nuc{7}{Li}$(\frac{3}{2}^-,\frac{1}{2})$ & -39.0(1)  & -39.24 & 2.28 [2.47]& 2.31(5) & 2.87  & 3.256  &-4.0(1)  &-4.00(3)
   \\
   \nuc{7}{Li}$(\frac{1}{2}^-,\frac{1}{2})$ & -38.9(1)  & -38.76 &
   \\
   \nuc{7}{Be}$(\frac{3}{2}^-,\frac{1}{2})$*& -37.4(1)  & -37.60 & 2.47 [2.28]& 2.51(2) &-1.06  &-1.398(15) &-6.7(1) 
   \\
   \nuc{7}{Be}$(\frac{1}{2}^-,\frac{1}{2})$*& -37.3(1)  & -37.17 &
   \\
   \nuc{8}{Li}$(2^+,1)$                     & -41.5(2)  & -41.28 & 2.10 [2.46]& 2.20(5) & 1.16(2)& 1.654 & 3.3(1) & 3.14(2)
   \\
   \nuc{8}{Li}$(2^+,1)$*                    & -41.0(2)  &        & 2.11 [2.48]&         & 1.13(3)&       & 3.0(4) &
   \\
   \nuc{8}{Li}$(1^+,1)$                     & -40.1(2)  & -40.30 &
   \\
   \nuc{8}{Li}$(3^+,1)$                     & -38.5(3)  & -39.02 &
   \\
   \nuc{8}{B}$(2^+,1)$                      & -37.5(2)  & -37.74 & 2.48 [2.11]&         & 1.45(1)& 1.036 & 5.9(4) & 6.83(21)
   \\
   \nuc{8}{B}$(2^+,1)$*                     & -37.8(2)  &        & 2.46 [2.10]&         & 1.42(2)&       & 6.5(2) &
   \\
   \nuc{8}{B}$(1^+,1)$*                     & -36.6(2)  & -36.97 &
   \\
   \nuc{8}{B}$(3^+,1)$*                     & -34.8(3)  & -35.42 &
   \\
   \nuc{9}{Be}$(\frac{3}{2}^-,\frac{1}{2})$ & -58.1(2)  & -58.16 & 2.37(1) [2.56(1)]& 2.38(1) &-1.18(1)&-1.178 & 5.1(1) & 5.29(4)
   \\
   \nuc{9}{Be}$(\frac{5}{2}^-,\frac{1}{2})$ & -55.7(2)  & -55.74 &
   \\
   \nuc{9}{B}$(\frac{3}{2}^-,\frac{1}{2})$* & -56.3(1)  & -56.31 & 2.56(1) [2.37(1)]&         & 2.97(1)&       & 4.0(3)
   \\
   \nuc{9}{B}$(\frac{5}{2}^-,\frac{1}{2})$* & -53.9(2)  & -53.95 &
   \\
   \nuc{9}{Li}$(\frac{3}{2}^-,\frac{3}{2})$ & -45.2(3)  & -45.34 & 1.97(1) [2.36(2)]& 2.11(5) & 2.66(3)& 3.437 &-2.3(1) & -3.06(2)
   \\
   \nuc{9}{Li}$(\frac{3}{2}^-,\frac{3}{2})$*& -45.9(3)  &        & 2.03(1) [2.45(1)]&         & 2.64(4)&       &-2.7(2) &
   \\
   \nuc{9}{Li}$(\frac{1}{2}^-,\frac{3}{2})$ & -43.2(4)  & -42.65 &
   \\
   \nuc{9}{C}$(\frac{3}{2}^-,\frac{3}{2})$  & -39.7(3)  & -39.04 & 2.45(1) [2.03(1)]&         &-0.75(3)&-1.391 &-4.1(4)
   \\
   \nuc{9}{C}$(\frac{3}{2}^-,\frac{3}{2})$* & -38.8(3)  &        & 2.36(2) [1.97(1)]&         &-0.82(4)&       &-3.7(1)
   \\
\end{tabular}
\end{ruledtabular}
\end{table*}
%
The energies $E$, point proton rms radii $r_p$ (and point neutron rms radii
$r_n$ for $N\neq Z$ nuclei), m.m.'s $\mu$ in IA, and quadrupole moments
$Q$ for the nuclear states calculated in this
work are presented in Table~\ref{tb:energies} along with experimental
values where available.
Experimental energies are from Ref.~\cite{Audi03}, EM moments are from 
Refs.~\cite{Purcell10,Tilley02,Tilley04,Borremans05}, and point radii are
converted from the charge radii given in Refs.~\cite{Mohr12,Amroun94,Shiner95,
Nortershauser11,Nortershauser09}.
Many energies for $A\le7$ nuclei evaluated with the AV18+IL7 Hamiltonian 
have been reported previously in Ref.~\cite{BPW11}.
The present energies, which are from independent calculations, are
in agreement with the previous results within the Monte Carlo statistical 
errors shown in parentheses.

For many of the isobaric analog states, the energy and moments are
calculated using the GFMC wave functions generated for the $T_z = -T$ state
and then simply interchanging protons and neutrons to evaluate the
$T_z = +T$ state.
These calculations are denoted by an asterisk (*) in the table,
and will be referred to as charge-symmetry-conserving (CSC) results.
For $^3$He, $^8$B, and $^9$C ground states we also made independent 
calculations with different starting VMC wave functions and different
isoscalar Coulomb terms~\cite{PPW07,GPK} in the GFMC propagator appropriate to the 
$T_z = +T$ state.
We then use these wave functions to predict the quantities in their
isobaric analogs, {\it i.e.}, $^3$H, $^8$Li, and $^9$Li.
The pairs of independent solutions for the isobaric
analogs will be referred to as charge-symmetry-breaking (CSB) results.
Thus six nuclear states have two entries in the table, comparing a direct 
calculation with the prediction by charge symmetry from its isobaric analog.

We see from the table that the energies in these paired calculations are
in generally good agreement, with the largest discrepancy for $A=9$,
where the difference is $\sim 2\%$ and the statistical errors almost touch.
In each of the CSB cases, the $T=-T_z$ state is more bound than its isobaric
analog, and the expectation values of individual terms in the nuclear 
Hamiltonian, like $\langle K_i \rangle$ and $\langle v_{ij} \rangle$, are
larger in magnitude.

The point nucleon rms radii are slightly larger for the proton-rich
nuclei compared to the charge symmetric solution from the proton-poor
isobaric analog. For example, the proton rms radius of $^9$Li ($^9$C)
is smaller (larger) when the appropriate Coulomb term is included in
the GFMC propagator, indicating that the system is more
compact (diffuse). If $^9$Li ($^9$C) is constructed from the $^9$C ($^9$Li)
solution, then it appears to be a more diffuse (compact) system.
Consistently with this weak sensitivity of the calculated
energies and radii to the charge symmetry picture implemented
to derive the nuclear wave functions, we find that the calculated
m.m.'s in IA are not statistically different in the $T=\frac{1}{2},1$ 
cases and we see only very weak evidence that the IA m.m.'s of the 
$A=9$, $T=3/2$ systems are affected by charge symmetry.
The quadrupole moments are also fairly consistent in the paired results
and close to the experimental values.

\subsection {Magnetic Moments in $A$=2--9 Nuclei}
\label{subsec:mm}
The calculations of the matrix elements, both diagonal and off diagonal,
have been described in detail in Refs.~\cite{PPW07,MPPSW08}. In particular,
the IA matrix element is evaluated using the $M1$ operator induced by
the one-body current given in Eq.~(\ref{eq:jlo}), namely
\begin{equation}
 {\bm \mu}^{\rm IA} =  \sum_i \left(e_{N,i}\, {\bf L}_i + \mu_{N,i}\,
  {\bm \sigma}_i\right)  \ .
\end{equation}
The matrix element associated with the MEC contribution is
\begin{eqnarray}
&&\langle J^\pi_f ,M_f\mid \! \mu^{\rm MEC}\! \mid J^\pi_i,M_i\rangle = \nonumber \\
&& -i \lim_{q\to 0} \frac{2\, m_N}{q}
\langle J^\pi_f,M_f\mid \! j_y^{\rm MEC}(q\, \hat{\bf x})\! \mid J^\pi_i,M_i\rangle \ ,
\end{eqnarray}
where the spin-quantization axis and momentum transfer ${\bf q}$ are, respectively,
along the $\hat{\bf z}$ and $\hat{\bf x}$ axes, and
$M_J=J$. The various contributions are evaluated for two small values
of $q<0.02$ fm$^{-1}$ and then extrapolated smoothly to the limit $q$=0.
The error due to extrapolation is much smaller than the statistical error
in the Monte Carlo sampling.

\begin{center}
\begin{table*}[tbh]
\caption{Magnetic moments in nuclear magnetons for $A=2$--$3$ nuclei evaluated
with SNPA and $\chi$EFT EM current operators. The current model labeled 
$\chi$EFT$^\dagger$ accounts for an exact calculation of the N2LO relativistic 
correction to the IA current (see text for explanation). Results labeled with 
HH are obtained utilizing trinucleon wave functions constructed with the 
hyperspherical harmonics (HH) method developed in 
Refs.~\cite{Kievsky97,Viviani06,Kievsky08}, and a nuclear Hamiltonian
consisting of AV18+UIX. The remaining results (except for the deuteron ones 
which are exact) are from GFMC calculations discussed in the text.
The spatial symmetry (s.s.) of the nuclear wave function is also given. 
The IS and IV labels indicate the isoscalar and isovector combinations.
Results obtained with the $\chi$EFT and $\chi$EFT$^\dagger$ models are not
predictions (see text for explanation).}
\label{tab:mm.2.3}
\begin{tabular}{c l c d d d d }
\hline
\hline
\\[1pt]
Nucleus($J^\pi$;$T$) & Current &
   \multicolumn{1}{c}{s.s.} &
   \multicolumn{1}{c}{IA} &
   \multicolumn{1}{c}{MEC} &
   \multicolumn{1}{c}{Total} &
   \multicolumn{1}{c}{Expt.}  \\ [1pt]
\hline
\\[1pt]
$n$($\frac{1}{2}^+$;$\frac{1}{2}$) & & & & & &-1.913 \\
$p$($\frac{1}{2}^+$;$\frac{1}{2}$) & & & & & & 2.793 \\[2pt]
IS                                 & & & & & & 0.440 \\
IV                                 & & & & & & 4.706 \\[2pt]
\hline
\\[1pt]
$^2$H($\frac{1}{2}^+$;$\frac{1}{2}$)
& SNPA                & [2]   &  0.8470 &  0.0012 &  0.8482     &  0.8574 \\
& $\chi$EFT           &       &  0.8470 &  0.0134 &  0.8604 (1) & \\
& $\chi$EFT$^\dagger$ &       &  0.8472 &  0.0102 &  0.8574     & \\ [2pt]
\hline
\\[1pt]
$^3$H($\frac{1}{2}^+$;$\frac{1}{2}$)
& SNPA                & [3]   &  2.556 (1) &  0.347 (2) &  2.903 (2) &  2.979 \\
& $\chi$EFT           &       &  2.556 (1) &  0.404 (1) &  2.960 (1) & \\ [1pt]
& $\chi$EFT$^\dagger$ (HH) &  &  2.569     &  0.410     &  2.979     & \\ [2pt]
$^3$He($\frac{1}{2}^+$;$\frac{1}{2}$)
&  SNPA               & [3]   & -1.743 (1) & -0.334 (2) & -2.077 (2) & -2.127 \\
& $\chi$EFT           &       & -1.743 (1) & -0.357 (1) & -2.100 (1) & \\
& $\chi$EFT$^\dagger$ (HH) &  & -1.749     & -0.378     & -2.127     & \\ [2pt]
IS     
& SNPA                &       &  0.407     &  0.006     &  0.413     &  0.426 \\
& $\chi$EFT           &       &  0.407     &  0.024     &  0.431     & \\
& $\chi$EFT$^\dagger$ (HH) &  &  0.410     &  0.016     &  0.426     & \\[2pt]
IV     
& SNPA                &       & -4.299     & -0.681     & -4.980     & -5.106 \\
& $\chi$EFT           &       & -4.299     & -0.761     & -5.060     & \\
& $\chi$EFT$^\dagger$ (HH) &  & -4.318     & -0.788     & -5.106     & \\[2pt]
\hline
\hline
\end{tabular}
\end{table*}
\end{center}
In Table~\ref{tab:mm.2.3}, we show, in addition to the proton and neutron experimental
m.m.'s, the experimental and calculated m.m.'s for the $A=2$ and $3$
nuclei, including MEC contributions from the EM currents in the
SNPA and $\chi$EFT models. 
In the table we label with IS and IV the isoscalar and isovector combinations
of the magnetic moments as given by:
\begin{equation}
\mu(T,T_z) = \mu({\rm IS}) + \mu({\rm IV}) T_z .
\end{equation}
With the label MEC we denote anything that goes
beyond the IA picture, therefore the $\chi$EFT MEC current includes also the
one-body relativistic correction operator entering at N2LO. The results for
the deuteron are from calculations
of matrix elements with wave functions which are exact solutions of the
two-body Schr\"odinger equation with the AV18 potential.
Results for the $A=3$ nuclei are from GFMC calculations with the nuclear Hamiltonian
consisting of the AV18 two-body and IL7 three-body potentials (AV18+IL7), while those
designated with HH are results from hyperspherical harmonics from Ref.~\cite{Piarulli12}
obtained with the nuclear Hamiltonian consisting of the AV18 and the
UIX three-body potentials (AV18+UIX).  Both the GFMC and HH wave functions
have been constructed separately without exploiting charge symmetry.
Strictly speaking, the GFMC m.m.'s are for the propagating Hamiltonian $H^\prime$,
{\it i.e.}, AV8$^\prime$+IL7$^\prime$, as discussed in Sec.~\ref{sec:qmc}.
The small $0.3\%$--$0.5\%$ difference between the GFMC and HH IA values
may be attributable to this difference in the Hamiltonians.

Numerical differences between the calculated $\chi$EFT MEC terms
are also affected by an additional approximation implemented in the GFMC calculations.
The one-body m.m.\ operator associated with the relativistic correction
at N2LO (illustrated in panel (d) of Fig.~\ref{fig:f1}) reads~\cite{Pastore08}
\begin{eqnarray}
{\bm \mu}^{\rm N2LO} =&& -\frac{e}{8\, m_N^3}\sum_{i=1}^A 
\Bigg[ \left\{ p_i^2 \, ,\, e_{N,i}\, {\bf L}_i + \mu_{N,i} \, {\bm \sigma}_i\right\} \nonumber \\
&&+ e_{N,i}\, {\bf p}_i\times({\bm \sigma}_i \times {\bf p}_i) \Bigg] \ ,
\label{eq:murc}
\end{eqnarray}
where ${\bf p}_i=-i\nabla_i$ and ${\bf L}_i$ are the linear momentum and
angular momentum operators of particle $i$, and $\{\dots\, ,\, \dots\}$ denotes
the anticommutator. In the GFMC calculations we do not explicitly evaluate
the ${\bf p}_i^2$ term, but instead approximate it with its average value, that is
${\bf p}_i^2\sim < {\bf p}_i^2> $, as determined from the expectation value
of the kinetic energy operator in each nucleus. This approximation leads,
in the case of $^3$He, to a $5\%$ difference in the MEC correction which itself
is a $20\%$ correction to the total calculated m.m., and to an even smaller
effect in $^3$H. The values utilized in the GFMC calculations are reported in
Table~\ref{tab:pav2}, for the nuclei investigated in this work. Of course,
this approximation only affects the N2LO $\chi$EFT operator.
The HH calculations use the non-approximated operator at
N2LO and are designated with $\chi$EFT$^\dagger$. These calculations
have been used to  constrain the LECs entering the $\chi$EFT currents,
therefore the HH results presented in Table~\ref{tab:mm.2.3} are not predictions,
in that they reproduce the experimental data by construction.
\begin{center}
\begin{table}[t]
\caption{Values (in units of fm$^{-2}$) of the approximated $\langle p_i^2\rangle$, entering the
$\chi$EFT current at N2LO, used in the GFMC calculations.} 
\label{tab:pav2}
\begin{tabular}{c | c c c c c c } 
\hline
\hline
$A$ & 2 & 3 & 6 & 7 & 8 & 9 \\
\hline
$\langle p_i^2\rangle$  & 0.5 & 0.8 & 1.2 & 1.2 & 1.2 & 1.3 \\
\hline
\hline
\end{tabular}
\end{table}
\end{center}

There are also tiny numerical differences between the SNPA calculations
presented here and those reported in Ref.~\cite{MPPSW08}. These may be
due to differences in the starting variational wave functions as well as
systematic uncertainties in the GFMC calculation. At any rate,
all the numerical differences mentioned above are of little importance
if one accounts for the sensitivity of these results to the nuclear
and current models utilized. The results in the $A=2$ and $3$ nuclei show
that the SNPA model underestimates the isoscalar component
in both the deuteron and the trinucleon m.m.'s.

\begin{center}
\begin{table*}[tbh]
\caption{Magnetic moments in nuclear magnetons for $A=6$--$9$ nuclei evaluated with
SNPA and $\chi$EFT EM current operators. Results labeled with a star
are obtained exploiting charge symmetry. The dominant spatial symmetry (s.s.)
of the nuclear wave function is given.}
\label{tab:mm.6.9}
\begin{tabular}{c l c d d d d }
\hline
\hline
Nucleus($J^\pi$;$T$) & Current &
   \multicolumn{1}{c}{s.s.} &
   \multicolumn{1}{c}{IA} &
   \multicolumn{1}{c}{MEC} &
   \multicolumn{1}{c}{Total} &
   \multicolumn{1}{c}{Expt.}  \\ [2pt]
\hline
$^6$Li($1^+$;$0$)
& SNPA      & [42] &  0.817 (1)& -0.010 (1)&  0.807 (1)&  0.822 \\
& $\chi$EFT &      &  0.817 (1)&  0.020 (1)&  0.837 (1)&        \\ [2pt]
\hline
$^7$Li($\frac{3}{2}^-$;$\frac{1}{2}$)
& SNPA      & [43] &  2.87 (1) &  0.25 (2) &  3.12 (2) &  3.256 \\
& $\chi$EFT &      &  2.87 (1) &  0.37 (1) &  3.24 (1) &        \\ [2pt]
$^7$Be($\frac{3}{2}^-$;$\frac{1}{2}$)*
& SNPA      & [43] & -1.06 (1) & -0.39 (2) & -1.45 (2) & -1.398 \\
& $\chi$EFT &      & -1.06 (1) & -0.36 (1) & -1.42 (1) &        \\ [2pt]
IS          
& SNPA      &      &  0.90     & -0.07     &  0.83     &  0.929 \\
& $\chi$EFT &      &  0.90     &  0.01     &  0.91     &        \\
IV          
& SNPA      &      & -3.93     & -0.64     & -4.57     & -4.654 \\
& $\chi$EFT &      & -3.93     & -0.73     & -4.66     &        \\
\hline
$^8$Li(2$^+$;1)
& SNPA &[431]      &  1.16 (2) &  0.20 (2) &  1.36 (3) &  1.654 \\
& $\chi$EFT &      &  1.16 (2) &  0.33 (1) &  1.49 (2) &        \\ [2pt]
$^8$B(2$^+$;1)*
& SNPA & [431]     &  1.42 (2) & -0.42 (2) &  1.00 (3) &  1.036 \\
& $\chi$EFT &      &  1.42 (2) & -0.31 (1) &  1.11 (2) &        \\ [2pt]
IS     
& SNPA      &      &  1.29     & -0.11     &  1.18     &  1.345 \\
& $\chi$EFT &      &  1.29     &  0.01     &  1.30     &        \\
IV     
& SNPA      &      &  0.13     & -0.31     & -0.18     & -0.309 \\
& $\chi$EFT &      &  0.13     & -0.32     & -0.19     &        \\
\hline
$^9$Li($\frac{3}{2}^-$;$\frac{3}{2}$)
& SNPA      & [432]&  2.66 (3) &  0.34 (4) &  3.00 (5) &  3.437 \\
& $\chi$EFT &      &  2.66 (3) &  0.70 (2) &  3.36 (4) &        \\ [2pt]
$^9$C($\frac{3}{2}^-$;$\frac{3}{2}$)
& SNPA      & [432]& -0.75 (3) & -0.48 (4) & -1.23 (5) & -1.391 \\
& $\chi$EFT &      & -0.75 (3) & -0.60 (3) & -1.35 (4) &        \\ [2pt]
IS
& SNPA      &      &  0.96     & -0.07     &  0.89     &  1.023 \\
& $\chi$EFT &      &  0.96     &  0.05     &  1.01     &        \\
IV
& SNPA      &      & -1.14     & -0.27     & -1.41     & -1.609 \\
& $\chi$EFT &      & -1.14     & -0.43     & -1.57     &        \\
\hline
$^9$Be($\frac{3}{2}^+$;$\frac{1}{2}$)
& SNPA      & [441]& -1.18 (1) & -0.12 (1) & -1.30 (1) & -1.178 \\
& $\chi$EFT &      & -1.18 (1) & -0.11 (1) & -1.29 (1) &        \\ [2pt]
$^9$B($\frac{3}{2}^+$;$\frac{1}{2}$)*
& SNPA      & [441]&  2.97 (1) & -0.10 (1) &  2.87 (1) &  \text {n.a.}  \\
& $\chi$EFT &      &  2.97 (1) &  0.09 (1) &  3.06 (1) &        \\ [2pt]
IS
& SNPA      &      &  0.89     & -0.11     &  0.78     & \\
& $\chi$EFT &      &  0.89     & -0.01     &  0.88     & \\
IV
& SNPA      &      &  4.15     &  0.02     &  4.17     & \\
& $\chi$EFT &      &  4.15     &  0.20     &  4.35     & \\
\hline
\hline
\end{tabular}
\end{table*}
\end{center}
In Table~\ref{tab:mm.6.9}, we report the GFMC calculations for the
m.m.'s of the $A=6$--$9$ nuclei. We compare results obtained using
either the SNPA or the $\chi$EFT MEC currents. The MEC corrections
evaluated in both models are qualitatively in agreement. They boost
the IA in the direction of the experimental data in all cases, except
for $^6$Li and $^9$Be. In these systems the IA results are already in 
very good agreement with the experimental data while the small MEC 
contributions make the predictions slightly worse.

The calculations denoted with an asterisk (*) are obtained exploiting
charge symmetry, that is by interchanging protons and neutrons to
generate the isobaric analogs wave functions.
For the $A=7$,$8$ nuclei, we verified that m.m.\ predictions obtained
with independent nuclear wave functions for the isobaric analogs,
{\it i.e.}, calculations which account for the proper isoscalar Coulomb term
in the starting VMC wave functions as well as in the GFMC propagator,
are essentially identical to the standard charge symmetric results, 
and therefore we do not report them.
In the $A=9, T=\frac{3}{2}$ case there is weak evidence for a CSB effect
so in this case we show the results of the two independent calculations.

Also in the $A=6$--$9$ nuclear m.m.'s, the difference between the SNPA
and $\chi$EFT corrections is more pronounced in the isoscalar component.
In all cases, the $\chi$EFT corrections are more positive (or less negative)
than the corresponding SNPA. This makes
the $\chi$EFT predictions closer to the experimental values. The isovector
corrections evaluated with the two models are reasonably in agreement with each other,
although they are bigger when derived from the $\chi$EFT model.
MEC corrections are crucial to bring the theory closer to the experimental values.
Their effect is particularly pronounced in the isovector combination of the $A=9$, $T=3/2$
nuclei's m.m.'s, for which the MEC SNPA ($\chi$EFT) correction provides $\sim20\%$
($\sim 30\%$) of the total calculated isovector contribution.

\begin{figure}
\includegraphics[height=0.35\textheight,angle=270,keepaspectratio=true]{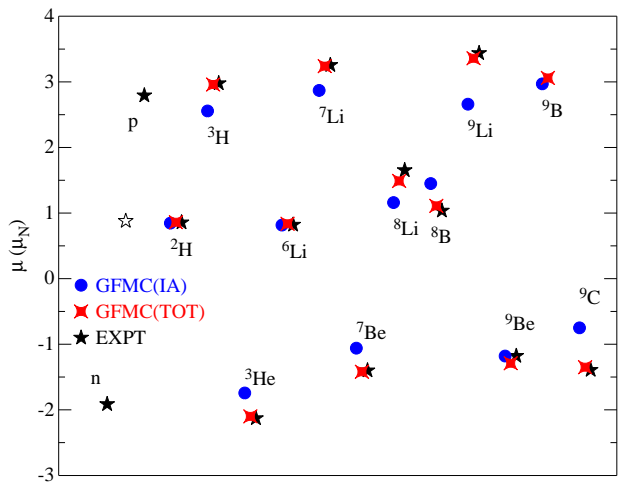}
\caption{(Color online)   Magnetic moments in nuclear magnetons for $A\leq9$ nuclei. 
           Black stars indicate
           the experimental values~\cite{Tilley02,Tilley04,Borremans05}, while blue dots (red diamonds)
           represent GFMC calculations which include the IA one-body EM current
           (total $\chi$EFT current up to N3LO). Predictions are for nuclei with $A>3$.
           }
\label{fig:f4}
\end{figure}

It is interesting to note that, despite the large effect observed in the 
$A=9$, $T=3/2$ systems, MEC corrections are considerably smaller in the $A=9$, $T=1/2$ 
nuclei.
This feature can be explained by considering the dominant spatial symmetry 
(s.s.)\ of the wave functions associated with the $A=9$ systems.
In particular, the dominant spatial symmetry of $^9$Be ($^9$B) is [441], 
corresponding to an $[\alpha, \alpha, n (p)]$ structure as shown in
Ref.~\cite{W06}.
A single nucleon outside an $\alpha$ particle feels no net OPE potential,
and this holds true also for a single nucleon outside a double-$\alpha$
[44] symmetry state.
Consequently, the NLO OPE currents illustrated in panels (b) and (c)
of Fig.~\ref{fig:f1}, which are generally the largest MEC terms in both
SNPA and $\chi$EFT approaches, do not contribute significantly.
On the other hand, the dominant spatial symmetry of $^9$C ($^9$Li) is
[432] $\sim$ [$\alpha$, $^3$He ($^3$H), $pp$ ($nn$)], and the NLO OPE term 
contributes in both the trinucleon clusters and in between the trinucleon 
clusters and the valence $pp$ ($nn$) pair.
The IA m.m.\ for $^9$Be is close to the experimental value, while
those for $^9$Li and $^9$C are far from the data, so this pattern of small
and large MEC corrections provides good overall agreement with the data.

The $\chi$EFT results reported in Tables~\ref{tab:mm.2.3} and ~\ref{tab:mm.6.9}
are summarized in Fig.~\ref{fig:f4}, where the experimental 
data~\cite{Purcell10,Tilley02,Tilley04,Borremans05}
(there are no data for the m.m.\ of $^9$B) are represented by black stars. 
We show also the experimental values for the proton and neutron m.m.'s, as 
well as their sum, which corresponds to the m.m.\ of an S-wave deuteron. 
The experimental values of the $A=2$--$3$ m.m.'s have been utilized to fix 
the LECs, therefore predictions are for $A>3$ nuclei.
The blue dots labeled as GFMC(IA) represent theoretical predictions obtained 
with the standard IA one-nucleon EM current entering at LO: diagram (a) of 
Fig.~\ref{fig:f1}.
The GFMC(IA) results reproduce the bulk properties of the m.m.'s of the light 
nuclei considered here. 
In particular, we can recognize three classes of nuclei with non-zero m.m.'s, 
{\it i.e.}, odd-even nuclei whose m.m.'s are driven by an unpaired valence proton, 
even-odd nuclei driven by an unpaired valence neutron, and odd-odd nuclei 
with either a deuteron cluster or a triton-neutron ($^3$He-proton) cluster
outside an even-even core.
Predictions which include all
the contributions to the N3LO $\chi$EFT EM currents illustrated in Fig.~\ref{fig:f1}
are represented by the red diamonds of Fig.~\ref{fig:f4}, labeled GFMC(TOT). 
In all cases except $^6$Li and $^9$Be (where the IA is already very good and
the MEC correction is very small) the predicted m.m.'s  are closer to the 
experimental data when the MEC corrections are added to the IA one-body EM 
operator.
\begin{figure*}
\includegraphics[height=6.0in,angle=270]{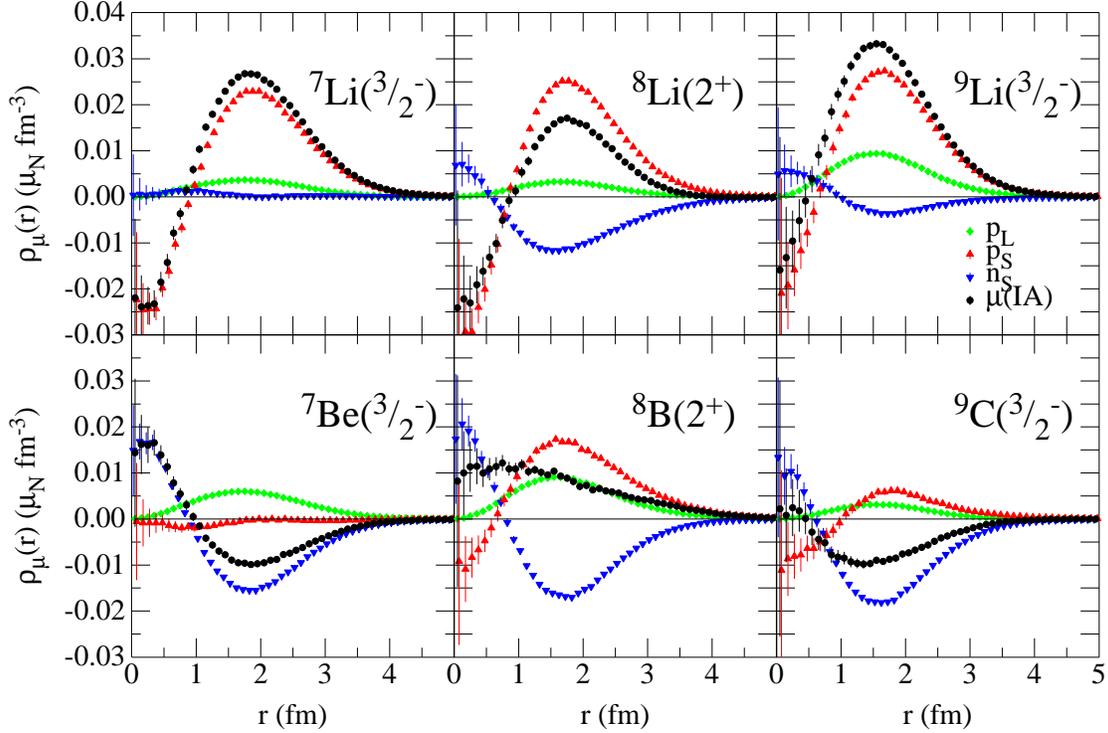}
\caption{(Color online)   Magnetic density in nuclear magnetons per fm$^3$ for 
selected nuclei (see text for explanation). }
\label{fig:mag_density}
\end{figure*}

It is also interesting to consider the spatial distribution of the various 
contributions to the m.m., {\it i.e.}, to examine the magnetic 
density.
The one-body IA contributions from the starting VMC wave functions are shown
in Fig.~\ref{fig:mag_density} for the isobaric analog pairs $^7$Li--$^7$Be,
$^8$Li--$^8$B, and $^9$Li--$^9$C.
(The VMC values for the IA m.m.'s are within a few \% of the final GFMC
values, so we expect their spatial distribution to be reasonably accurate.)
In the figure, the red upward-pointing triangles are the contribution from
the proton spin, $\mu_p[\rho_{p\uparrow}(r) - \rho_{p\downarrow}(r)]$, 
and similarly the blue downward-pointing triangles are the contribution 
from the neutron spin.
The green diamonds are the proton orbital (convection current)
contribution, and the black circles are the sum.
The integrals of the black curves over $d^3r$ give the total m.m.'s
of the nuclei in IA.

For the neutron-rich lithium isotopes, there is one unpaired proton (embedded
in a $p$-shell triton cluster) with essentially the same large positive 
contribution in all three cases.  
The proton orbital term is also everywhere positive, but relatively small.
For $^7$Li and $^9$Li, the neutrons are paired up, and give only a small 
contribution, so the total m.m.\ is close to the sum of the
proton spin and orbital parts.  However $^8$Li has one unpaired neutron which
acts against the proton and significantly reduces the overall m.m.\ values.
For the proton-rich isobaric analogs, there is one unpaired neutron (embedded
in a $p$-shell $^3$He cluster) with the same sizable negative contribution 
in all three cases. 
In $^7$Be and $^9$C, the protons are paired up and give little net 
contribution, but the orbital term is always positive and acts against
the neutron spin term.
In $^8$B there is also one unpaired proton, which gives a bigger contribution
than the unpaired neutron and results in a net positive m.m.\ value.

\begin{table*}[tbh]
\begin{center}
\caption{Magnetic moments in nuclear magnetons of $A \leq 9$ nuclei evaluated with
the $\chi$EFT current operator. Individual contributions entering at
LO (IA), NLO, N2LO, N3LO are shown (see text for explanation). Remaining notation
is as defined in Table~\ref{tab:mm.6.9}.}
\label{tab:break}
\footnotesize
\begin{tabular}{c l l l l l l c l l l l l }
\hline
\hline
\\
Nucleus($J^\pi$;$T$) & \multicolumn{1}{c}{s.s.} &\multicolumn{1}{c}{IA}
                     & \multicolumn{1}{c}{NLO}  &\multicolumn{1}{c}{N2LO}
                     & \multicolumn{2}{c}{N3LO} &\multicolumn{2}{c}{N3LO-LECs}
                     & \multicolumn{1}{c}{MEC}  &\multicolumn{1}{c}{Total}
                     & \multicolumn{1}{c}{Expt.} \\
\cline{6-7} \cline{8-9}
\vspace{-0.1in}
\\
& & & & &  \multicolumn{1}{c}{LOOP} & \multicolumn{1}{c}{MIN}
&   \multicolumn{1}{c}{NM}  & \multicolumn{1}{c}{OPE} &  &   & 
\vspace{ 0.1in}
\\
\hline
\vspace{-0.1in}
\\
$^2$H($1^+$;$0$) &[2] &{\color{white}-}0.8470    & {\color{white}-}---
                      &               -0.0042    & {\color{white}-}---
                      &{\color{white}-}0.0364   
                      &               -0.0135    &                -0.0052
                      &{\color{white}-}0.0135   
                      &{\color{white}-}0.8604    &{\color{white}-}0.8574\\ [6pt]
\hline
\vspace{-0.1in}
\\
$^3$H($\frac{1}{2}^+$;$\frac{1}{2}$)& [3]
                    &{\color{white}-}2.556 (1)     &{\color{white}-}0.253
		    & -0.033                       &{\color{white}-}0.058 
                    &{\color{white}-}0.035 
                    & -0.011              &{\color{white}-}0.102 
                    &{\color{white}-}0.404
                    &{\color{white}-}2.960 (1) & {\color{white}-}2.979 \\ [6pt]
$^3$He($\frac{1}{2}^+$;$\frac{1}{2}$)& [3] 
               &-1.743 (1)     &-0.248           & {\color{white}-}0.024 
               & -0.055    & {\color{white}-}0.056   & -0.022  
               & -0.110               & -0.357        &-2.100 (1)   & -2.127\\ [6pt]
IS                 & &\,\,0.407 (1)      &\,\,0.002   &  -0.005         &\,\,0.001
                     &\,\,0.046   &-0.017     &-0.004
                     &\,\,0.023   &\,\,0.430 (1)  &\,\,0.426\\[6pt]
IV                 & &\,\,4.299 (1)     & -0.501    &\,\,0.057      &-0.113 
                     &\,\,0.020    & -0.011    & -0.213
                     &  -0.760      &-5.060 (1)    &-5.106  \\[6pt]
\hline
\vspace{-0.1in}
\\
$^6$Li($1^+$;$0$)&[42]&{\color{white}-}0.817 (1) & {\color{white}-}---
                      &               -0.012     & {\color{white}-}---
                      &{\color{white}-}0.063
                      &               -0.023     &                -0.007
                      &{\color{white}-}0.020 (1)
                      &{\color{white}-}0.837 (1) &{\color{white}-}0.822\\ [6pt]
\hline
\vspace{-0.1in}
\\
$^7$Li($\frac{3}{2}^-$;$\frac{1}{2}$)& [43]
                    &{\color{white}-}2.87 (1)  &{\color{white}-}0.237
		    &               -0.062     &{\color{white}-}0.064
                    &{\color{white}-}0.034
                    &               -0.012     &{\color{white}-}0.107
                    &{\color{white}-}0.37 (1)
                    &{\color{white}-}3.24 (1)& {\color{white}-}3.256 \\ [6pt]
$^7$Be($\frac{3}{2}^-$;$\frac{1}{2}$)*& [43]
               &-1.06 (1)        & -0.237         &{\color{white}-}0.015
               &-0.064           & {\color{white}-}0.072  & -0.027
               &-0.120           &-0.36 (1)      &-1.42 (1)& -1.398\\ [6pt]
IS                 & &\,\,0.90    &\,\,---   &   -0.024      &\,\,---
                     &\,\,0.053   &-0.020    &   -0.006
                     &\,\,0.01    &\,\,0.91 &\,\,0.929\\[6pt]
IV                 & &-3.93       &-0.473    &\,\,0.078    &-0.127
                     &\,\,0.038   &-0.014    &-0.227
                     &-0.73       &-4.66    &-4.654  \\[6pt]
\hline
\vspace{-0.1in}
\\
$^8$Li(2$^+$;1)& [431]& {\color{white}-}1.16 (2)  &{\color{white}-}0.226
                      &                -0.038     &{\color{white}-}0.045
                      & {\color{white}-}0.056     &               -0.021
                      & {\color{white}-}0.062     &{\color{white}-}0.33 (1)
                      & {\color{white}-}1.49 (2)  &{\color{white}-}1.654 \\ [6pt]
$^8$B(2$^+$;1)* &[431]& {\color{white}-}1.42 (2)  &               -0.226
                      &                -0.020     &               -0.045
                      & {\color{white}-}0.090     &               -0.033
                      &                -0.077     &-0.31 (1)
                      & {\color{white}-}1.11 (2)  &{\color{white}-}1.036  \\ [6pt]

IS               & &\,\,1.29  & \,\,---
                   &   -0.029 & \,\,---
                   &\,\,0.073 &-0.027
                   & -0.007   &\,\,0.01
                   &\,\,1.30  &\,\,1.345      \\[6pt]
IV               & &\,\,0.13  &-0.226
                   &\,\,0.009 &-0.045
                   &\,\,0.017 &-0.006
                   &   -0.070 &-0.32
                   &   -0.19  &-0.309   \\[6pt]
\hline
\vspace{-0.1in}
\\
$^9$Li($\frac{3}{2}^-$;$\frac{3}{2}$)
             &[432]   & {\color{white}-}2.66 (3)  &{\color{white}-}0.403
                      &                -0.076     &{\color{white}-}0.141
                      & {\color{white}-}0.108     &               -0.016
                      & {\color{white}-}0.141     &{\color{white}-}0.70 (2)
                      & {\color{white}-}3.36 (4)  &{\color{white}-}3.437 \\ [6pt]
$^9$C($\frac{3}{2}^-$;$\frac{3}{2}$)
             &[432]   &                -0.75 (3)  &               -0.372
                      & {\color{white}-}0.039     &               -0.135
                      & {\color{white}-}0.058     &               -0.031
                      &                -0.156     &               -0.60 (3)
                      &                -1.35 (4)  &               -1.391\\ [6pt]
IS       &   & \,\,0.96 & \,\,0.015
             &    -0.019 & \,\,0.003
             & \,\,0.083 &    -0.024
             &    -0.008 & \,\,0.05
             & \,\,1.01 & \,\,1.023 \\[6pt]
IV       &   &   -1.14  & -0.258
             &   \,\,0.038  & -0.092
             & -0.017 & -0.005
             & -0.099 &   -0.43
             &    -1.57 &   -1.609 \\[6pt]
\hline
\vspace{-0.1in}
\\
$^9$Be($\frac{3}{2}^+$;$\frac{1}{2}$)
& [441] & -1.18 (1)   &                -0.084
                      & {\color{white}-}0.019     & -0.037
                      & {\color{white}-}0.041     & -0.009
                      &                -0.038     & -0.11 (1)
                      &                -1.29 (1)  & -1.178 \\ [6pt]
$^9$B($\frac{3}{2}^+$;$\frac{1}{2}$)*
& [441] &{\color{white}-}2.97 (1)                 &{\color{white}-}0.084
                      &           -0.057          &{\color{white}-}0.037
                      &           -0.005          &               -0.004
                      & {\color{white}-}0.032     &{\color{white}-}0.09 (1)
                      & {\color{white}-}3.06 (1)&\,\,\,{\rm n.a.} \\[6pt]
IS      & &\,\,0.89   & {\color{white}-}---
                      &                -0.019      &\,\,---
                      &              \,\,0.018     &-0.007
                      &                 -0.003     &-0.01
                      & \,\,0.88\\[6pt]
IV      & &\,\,4.15   & \,\,0.168
                      & -0.076          &\,\,0.074
                      & -0.046          &\,\,0.005
                      &  \,\,0.070          &\,\,0.20
                      &\,\,4.35\\[6pt]
\hline
\hline
\end{tabular}
\end{center}
\end{table*}

In Table~\ref{tab:break}, we explicitly show the various contributions entering the $\chi$EFT
operator. The labeling in the table has been defined in Sec.~\ref{sec:chieft}.
We list the contributions at each order. At N3LO, we separate the terms that do not
depend on EM LECs ({\it i.e.} the LOOP contribution and the contact MIN currents; the former
depends on the known axial coupling constant, $g_A$, and pion decay amplitude, $F_\pi$,
while the latter depends on the strong LECs entering the $N\!N$
$\chi$EFT potential at N2LO) and those that depend on them ({\it i.e.} the contact
NM and the OPE current whose isovector component has been saturated with the $\Delta$
transition current). In most cases, chiral convergence is observed but for
the isovector N3LO OPE contribution whose order of magnitude is in some cases comparable to the
OPE contribution at NLO. It is likely that the explicit inclusion of $\Delta$ degrees
of freedom in the present $\chi$EFT would significantly
improve the convergence pattern, since in such a theory
this isovector OPE current, presently entering at N3LO, would be promoted to N2LO.

In Table~\ref{tab:break}, we do not provide the errors associated with the individual terms
at each order because they are highly correlated. We limit ourselves to report the
errors associated with the IA, MEC, and total results.
Also in this table, we denote calculations performed enforcing charge symmetry
with an asterisk (*). In these calculations, the isoscalar component of pure
isovector operators (that is the OPE operator at NLO and the LOOP operator at N3LO)
are obviously zero. Calculations in which the nuclear wave functions are constructed
independently present an isoscalar admixture even in the purely isovector corrections,
namely the NLO and N3LO LOOP.
We do not report the individual contributions entering the SNPA calculations. For
the $A\leq 7$ nuclei, they are found to be in agreement with those reported
in Ref.~\cite{MPPSW08}.

\begin{table}
\begin{center}
\caption{Magnetic moments in nuclear magnetons of the $A=9$, $T= 3/2$ systems.
The nuclear wave functions are derived within a CSB picture
(first block), and in a CSC framework (last two blocks).
See text for further explanations.}
\label{tab:mu9}
\squeezetable
\begin{tabular}{l d d d d}
\hline
\hline
\\[2pt]
 Nucleus($J^\pi$;$T$)&
   \multicolumn{1}{c}{IA} &
   \multicolumn{1}{c}{MEC} &
   \multicolumn{1}{c}{Total} &
   \multicolumn{1}{c}{Expt.}  \\ [2pt]
\hline
\\
$^9$Li($\frac{3}{2}^-$;$\frac{3}{2}$)
         &  2.66 (3) &  0.70 (2) &  3.36 (4) &  3.437 \\
$^9$C($\frac{3}{2}^-$;$\frac{3}{2}$)
         & -0.75 (3) & -0.60 (3) & -1.35 (4) & -1.391 \\
IS      
         &  0.96     &  0.05     &  1.01     &  1.023 \\
IV     
         & -1.14     & -0.43     & -1.57     & -1.609 \\
\hline
\\
$^9$Li($\frac{3}{2}^-$;$\frac{3}{2}$)
         &  2.66 (3) &  0.70 (2) &  3.36 (4) &  3.437 \\
$^9$C($\frac{3}{2}^-$;$\frac{3}{2}$)*
         & -0.82 (4) & -0.68 (2) & -1.50 (4) & -1.391 \\
IS
         &  0.93     &  0.01     &  0.94     &  1.023 \\
IV
         & -1.16     & -0.46     & -1.62     & -1.609 \\
\hline
\\
$^9$Li($\frac{3}{2}^-$;$\frac{3}{2}$)*
         &  2.64 (4) &  0.62 (3) &  3.26 (5) &  3.437 \\
$^9$C($\frac{3}{2}^-$;$\frac{3}{2}$)
         & -0.75 (3) & -0.60 (3) & -1.35 (4) & -1.391 \\
IS
         &  0.94     &  0.01     &  0.95     &  1.023 \\
IV
         & -1.13     & -0.40     & -1.53     & -1.609 \\[6pt]
\hline
\hline
\end{tabular}
\end{center}
\end{table}
\begin{table}
\begin{center}
\caption{Spin expectation value $\left< \sigma_z \right>$ for the mirror nuclei
$^9$Li and $^9$C evaluated within CSB and
CSC frameworks. (See text for explanation).}
\label{tab:sigma_final}
\begin{tabular}{ l | l l l }
\hline
\hline
   \multicolumn{1}{c}{Symmetry} &
   \multicolumn{1}{c}{IA} &
   \multicolumn{1}{c}{Total} &
   \multicolumn{1}{c}{Expt.}  \\ [2pt]
\hline
CSB : $^9$Li($\frac{3}{2}^-$;$\frac{3}{2}$),\,\,\,$^9$C($\frac{3}{2}^-$;$\frac{3}{2}$)  & 1.11 (11)  & 1.37 (15) & 1.44 \\
CSC : $^9$Li($\frac{3}{2}^-$;$\frac{3}{2}$),\,\,\,$^9$C($\frac{3}{2}^-$;$\frac{3}{2}$)* & 0.95 (1) & 1.00 (1) & \\
CSC : $^9$Li($\frac{3}{2}^-$;$\frac{3}{2}$)*,$^9$C($\frac{3}{2}^-$;$\frac{3}{2}$)& 1.00 (1) & 1.05 (1) & \\[6pt]
\hline
\hline
\end{tabular}
\end{center}
\end{table}

The $A=9$, $T=3/2$ nuclei are very interesting systems not only for the large 
effect produced by the isovector MEC correction, but also for the `anomaly' 
associated with their isoscalar component. 
In Ref.~\cite{Utsuno04} and references therein, the role of mirror 
symmetry---or charge symmetry---in the $A=9$, $T=3/2$ nuclei has been 
investigated. 
In particular, if mirror symmetry is assumed, that is if the equality of 
$nn$ and $pp$ forces is enforced, then the isoscalar spin expectation value 
$<\sigma_z>$ can be deduced as
\begin{equation}
\label{eq:sigmaz}
 <\sigma_z> = \frac{\mu(T_z=+T) + \mu(T_z=-T) - J}{ (g_s^p +g_s^n -1)/2} =
\frac{2 \mu ({\rm IS}) - J}{0.3796} \ .
\end{equation}
In the equation above, $\mu$ and $\mu({\rm IS})$ are the total m.m. and
the isoscalar combination of the mirror nuclei m.m.'s, $J$ is the total angular
momentum and $g_s^{p(n)}$ is the spin $g$ factor of the proton (neutron).
The experimental value of $<\sigma_z>$ is $1.44$~\cite{Matsuta95} which
is---quoting from Ref.~\cite{Utsuno04}---``anomalously large
if compared to the single-particle estimate of 1.''

Driven by the discussion reported in Ref.~\cite{Utsuno04},
we calculate the $<\sigma_z>$ value for the $A=9$, $T=3/2$ nuclei using the
CSB and CSC models described at the beginning of this section.
To make this last comment explicit, we show three different calculations
for the $^9$Li and $^9$C m.m.'s in Table~\ref{tab:mu9} using the $\chi$EFT MEC.
The first calculation is performed by constructing the nuclear wave functions
independently including the appropriate Coulomb term in the
GFMC propagator for each nucleus.
The second (third) CSC calculation uses a $^9$C ($^9$Li) wave function which
is constructed from that of $^9$Li ($^9$C) by interchanging protons with 
neutrons, {\it i.e.}, by imposing mirror symmetry.
The CSB isoscalar component is obtained combining the m.m.'s of $^9$C and $^9$Li
obtained independently. Therefore the error associated with this observable
follows from propagating the statistical errors of the calculated
m.m.'s of the $A=9$, $T=3/2$ nuclei.
Within the CSC picture, the error associated with the isoscalar
combination of the $A=9$, $T=3/2$ m.m.'s can be directly evaluated in
the GFMC calculation. These CSC isoscalar values are found to be very stable,
with a statistical error of less that $\sim 1\%$.

We use the CSC and CSB IA and total values for the isoscalar combination
of the $^9$C and $^9$Li m.m.'s to evaluate the $<\sigma_z>$ value
as given in Eq.~(\ref{eq:sigmaz}). The results are shown in
Table~\ref{tab:sigma_final}. From this Table, we see that the
CSC values are consistent with the single particle prediction of
$1$. We observe that, in the CSB calculation both the
isoscalar IA and MEC corrections are larger than those obtained
in the mirror symmetry based picture. However, the error associated with
this last calculation does not allow us conclude with certainty
that implementing a CSB picture would resolve the aforementioned `anomaly'
associated with the $A=9$, $T=3/2$ nuclei.
Within the statistics available at present, we can argue that, in order
to reproduce the experimental $<\sigma_z>$ value, one would have to
combine both the effect of CSB and the correction due to the MEC
currents.
As briefly mentioned before, we investigated the
role of mirror symmetry breaking also in the $A=7,8$ nuclei and found that the
calculated m.m.'s are not sensitive to the different nuclear wave functions.
A trivial argument to justify this result is to be found in the isospin of the
investigated nuclei: intuitively, one would expect the effect of mirror 
symmetry breaking to increase as $\Delta T_z = 2 \, T$ of the mirror nuclei
becomes larger.

\subsection {Electromagnetic Transitions in $A$=6--9 Nuclei}
\label{subsec:trans}
In Table~\ref{tab:finalm1tra}, we report the results for the $M1$ transition 
matrix elements and the transition widths in nuclei with mass number $A\leq 9$.
For these calculations---obtained with the $\chi$EFT EM current operator---we
report only the IA and the MEC contributions.
Experimental data in the table are taken from Refs.~\cite{Tilley02,Tilley04}.
The widths $\Gamma$ in units of MeV are calculated as
\begin{eqnarray}
\label{eq:gammaM}
 \Gamma_{M1} = 0.890 \left (\frac{\Delta E}{\hbar c} \right)^3 B(M1) \ ,
\end{eqnarray}
where $\hbar c$ is in units of MeV~fm, and $B(M1)$ is the squared reduced 
matrix element of the m.m.\ operator between the initial and the final 
nuclear state divided by $(2 J_i+1)$, with $J_i$ the initial state
angular momentum. In the equation above, $\Delta E$ is the energy difference
between the final and the initial state (in units of MeV) for which we take the
experimental values as given in Refs.~\cite{Tilley02,Tilley04}.

These calculations are obtained, as before, by propagating up to $\tau = 0.8$ MeV$^{-1}$
with an evaluation after every 40 propagation steps, {\it i.e.}, at intervals of
$\tau = 0.02$ MeV$^{-1}$.  The analysis of the
IA and the MEC contributions are performed separately in the same fashion that has
been implemented for the m.m.'s.

The predictions for the $A=6,7$ nuclei as well as those for the $A=8$, 
($1^+\!\rightarrow\!2^+$)
transitions are in very good agreement with the experimental data. In all these
cases the MEC corrections are needed to bring the theory in agreement with the
experimental data. Results for the ($3^+\rightarrow 2^+$) transitions in the
$A=8$ systems underpredict the experimental data, however the latter have large
experimental errors, and thus it is difficult to reach any robust conclusions
as to the actual level of agreement between theory and experiment.
The transition in $^9$Be is known with good accuracy, but the predicted width
is lower than the experimental data although the error bars almost touch.
We also report a prediction for the 
($\frac{1}{2}^-\!\rightarrow\!\frac{3}{2}^-$) $M1$ transition 
in $^9$Li which has not been measured yet.
We did not calculate the $^9$C transition to its unbound $\frac{1}{2}^-$
state.

\begin{table*}
\begin{center}
\caption{Matrix elements in units of nuclear magnetons and widths of $M1$ transitions in $A=6$--$9$ nuclei which account for
the $\chi$EFT current operator up to N3LO. IA and MEC contributions are shown.}
\label{tab:finalm1tra}
\begin{tabular}{c  c  l l l l }
\hline
\hline
\\[2pt]
$(J^\pi_i\rightarrow J^\pi_f)$ & $M1$ and $\Gamma$ &
   \multicolumn{1}{c}{IA} &
   \multicolumn{1}{c}{MEC} &
   \multicolumn{1}{c}{Total} &
   \multicolumn{1}{c}{Expt.}  \\ [2pt]
\hline
\\[2pt]
$^6$Li($0^+\rightarrow 1^+$)
& M1                       &    3.63 (1) & 0.38  & 4.01 (1)& \\
& $\Gamma$ (eV)&   6.90 (2)&  & 8.41 (3) & 8.19 (17)\\[6pt]
\hline
\\[2pt]
$^7$Li($\frac{1}{2}^-\rightarrow \frac{3}{2}^-$)
& M1                      & 2.66 (1) & 0.47 (1) & 3.13 (2) & \\
& $\Gamma$ ($10^{-3}$ eV) & 4.47 (5) &          & 6.19 (8) & 6.30 (31)\\[6pt]
$^7$Be($\frac{1}{2}^-\rightarrow \frac{3}{2}^-$)
& M1                      & 2.31 (2) & 0.41 (1) & 2.72 (2)&\\ 
& $\Gamma$ ($10^{-3}$ eV) & 2.44 (4)  &  & 3.39 (6) &3.43 (45)\\[6pt]
\hline
\\[2pt]
$^8$Li($1^+\rightarrow 2^+$)
& M1                     & 3.47 (4)& 0.74 (2) &  4.21 (5)& \\
& $\Gamma$ ($10^{-2}$ eV)& 4.4 (1)& & 6.5 (2) & 5.5 (1.8)\\[6pt]
$^8$B($1^+\rightarrow 2^+$)
& M1                     & 3.17 (5)& 0.67 (2)& 3.84 (6)&\\
& $\Gamma$ ($10^{-2}$ eV)& 1.8  (1)&         & 2.6 (1)& 2.52 (11)\\[6pt]
\hline
\\[1pt]
$^8$Li($3^+\rightarrow 2^+$)
& M1                     & 0.98 (6)  & 0.20 (5)  &  1.17 (8)& \\
& $\Gamma$ ($10^{-2}$ eV)&  1.8 (2)& &2.6 (3)&  7.0 (3.0)\\[6pt]
$^8$B($3^+\rightarrow 2^+$)
& M1                     & 1.31 (6)& 0.23 (5) & 1.56 (8)&\\
& $\Gamma$ ($10^{-2}$ eV)& 3.5 (3)&        &  4.9 (5)& 10 (5) \\[6pt]
\hline
\\[2pt]
$^9$Li($\frac{1}{2}^-\rightarrow \frac{3}{2}^-$)
& M1                     &  2.28 (3)& 0.36 (4)&2.64 (5)& \\
& $\Gamma$ ($10^{-1}$ eV)&  5.9 (2)& &  7.9 (3)& n.a.\\[6pt]
\hline
\\[2pt]
$^9$Be($\frac{5}{2}^-\rightarrow \frac{3}{2}^-$)
& M1                     &  1.42 (3) & 0.20 (2)&1.62 (4)& \\
& $\Gamma$ ($10^{-2}$ eV)&  5.6 (3) & & 7.2 (4)& 8.9 (1.0)\\[6pt]
\hline
\hline
\end{tabular}
\end{center}
\end{table*}

\begin{figure*}
\includegraphics[height=6.0in,angle=270]{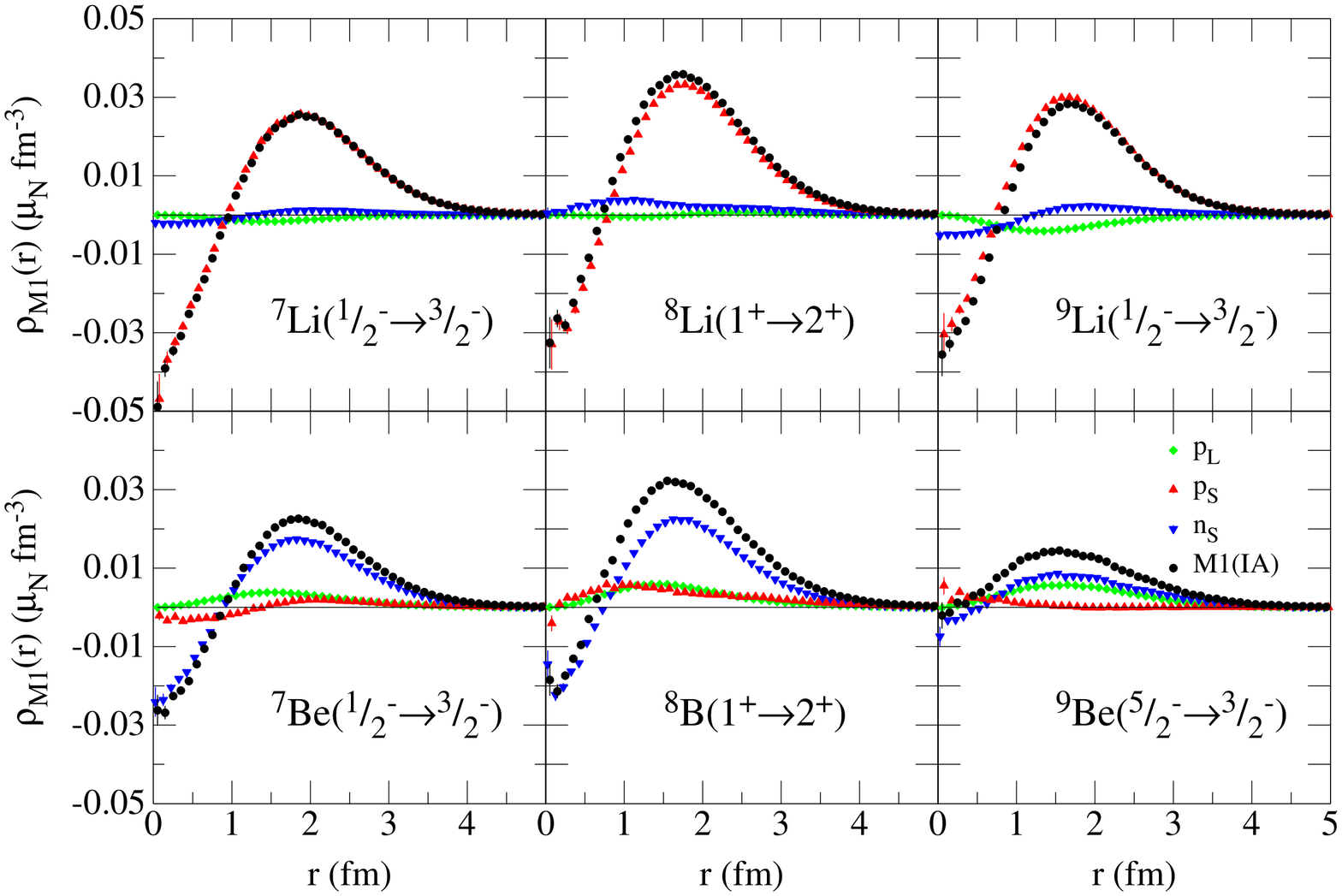}
\caption{(Color online) $M1$ transition density in nuclear magnetons per fm$^3$ 
for selected nuclei (see text for explanation). }
\label{fig:m1_density}
\end{figure*}

The magnetic transition densities in IA as obtained from the VMC starting wave
functions are shown in Fig.~\ref{fig:m1_density}.
As before, the red upward-pointing triangles are the contribution from the
proton spin term, the blue downward-pointing triangles are from the neutron
spin, the green diamonds are from the proton orbital term, and the black 
circles are the total IA contribution.
For the lithium isotopes, the transitions are predominantly from the proton
spin term, {\it i.e.}, these are almost pure proton spin-flip transitions.
For $^7$Be and $^8$B, the neutron spin term is the most important, but with
some contribution from the proton spin and orbital terms.
The neutron spin-flip is also the biggest term in the $^9$Be transition,
but here the proton orbital piece is almost the same size.

Finally in Table~\ref{tab:final2tra}, we show IA results for the electric 
quadrupole matrix elements and the associated transition widths. 
The latter in units of MeV is
\begin{eqnarray}
\label{eq:gammaE}
 \Gamma_{E2} = 0.241 \left (\frac{\Delta E}{\hbar c} \right)^5 B(E2) \ ,
\end{eqnarray}
where $B(E2)$ is the square of the reduced matrix element of the electric quadrupole 
operator given by
\begin{equation} 
 \rho^{\rm IA} = \sum_i e_{N,i}\, r_i^2 \, Y_2(\hat{\bf r}_i) \ ,
\end{equation}
where $Y_L$ is the spherical harmonic.
The IA picture provides a decent description of the two experimental data
points that are available, which might possibly be improved by the inclusion
of two-body effects. This topic has not been addressed in this work
although effort in this direction is underway.

\begin{figure}[h!]
\includegraphics[height=0.45\textheight,keepaspectratio=true]{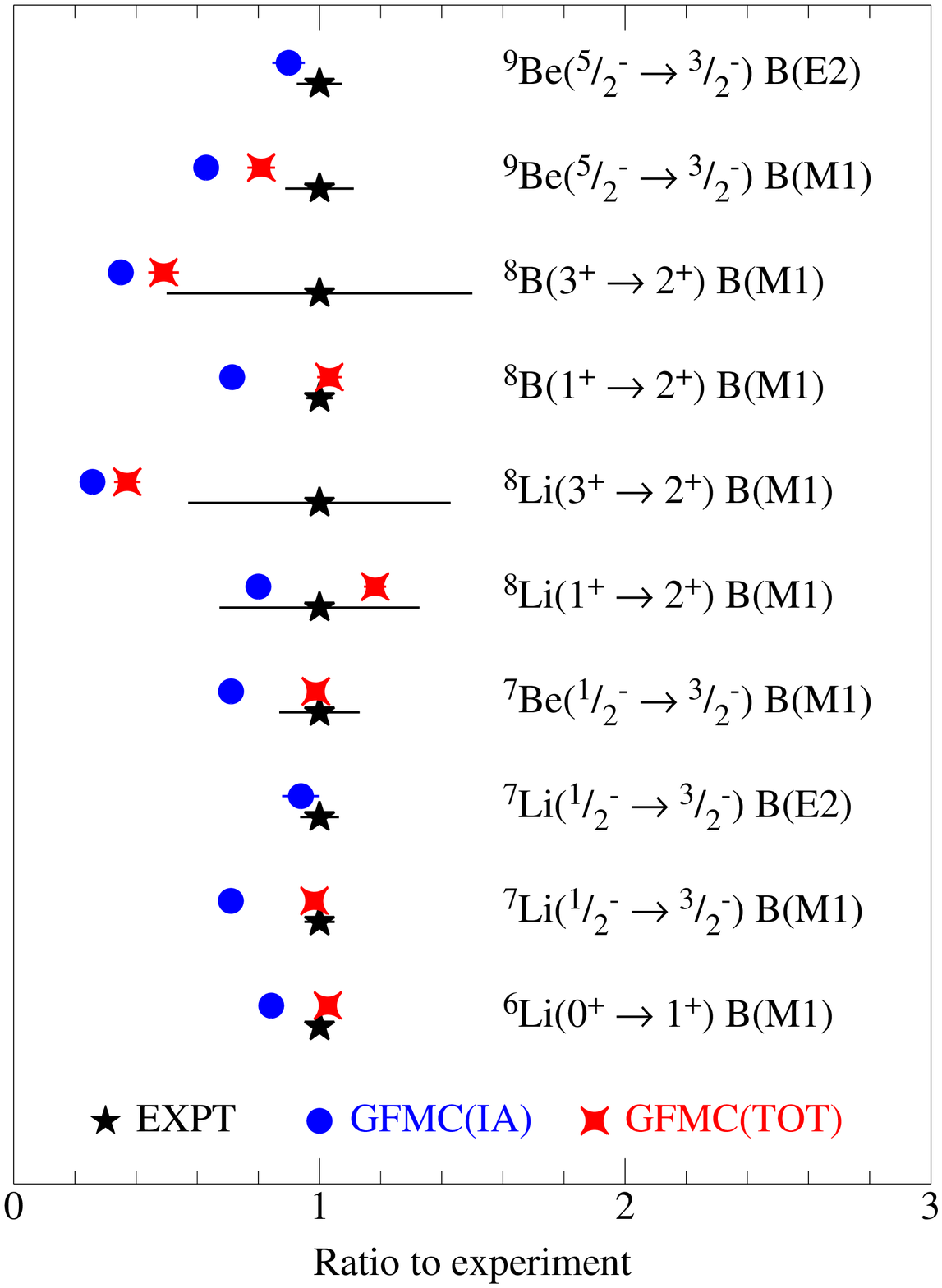}
\caption{(Color online)   Ratio to the experimental $M1$ and $E2$ transition widths in $A\leq9$ nuclei.
           Black stars with error bars indicate the experimental 
           values~\cite{Tilley02,Tilley04}, while blue dots (red diamonds)
           represent GFMC calculations which include the IA one-body EM current
           (total $\chi$EFT current up to N3LO).
           }
\label{fig:f5}
\end{figure}

The results discussed in this section are summarized in Fig.~\ref{fig:f5}
for EM transitions whose widths are known experimentally.
We observe that in most cases, the agreement with the experimental data
is excellent, and that the MEC contributions are crucial for the $B(M1)$ cases.

\begin{table}
\begin{center}
\caption{Matrix elements in units of $e$ fm$^2$ and widths of $E2$
transitions in $A=7$--$9$ nuclei. Only IA results are shown.}
\label{tab:final2tra}
\begin{tabular}{c  c  r r}
\hline
\hline
\\[2pt]
$(J^\pi_i\rightarrow J^\pi_f)$ & $E2$ and $\Gamma$ & IA  & Expt.   \\ [2pt]
\hline
\\[2pt]
$^7$Li($\frac{1}{2}^-\rightarrow \frac{3}{2}^-$) & E2        & 5.59 (16) &  \\
& $\Gamma$ ($10^{-7}$ eV)                        & 3.1 (2)   & 3.30 (21)    \\[6pt]
$^7$Be($\frac{1}{2}^-\rightarrow \frac{3}{2}^-$) & E2        & 9.43 (24)&  \\
& $\Gamma$ ($10^{-7}$ eV)                        & 5.2 (3)   & {\rm n.a.} \\[6pt]
\hline
\\[2pt]
$^8$Li($1^+\rightarrow 2^+$)& E2      & 2.04 (8)  & \\
& $\Gamma$ ($10^{-6}$ eV)   & 1.0 (1)&    {\rm n.a.}  \\[6pt]
$^8$B($1^+\rightarrow 2^+$) & E2      & 4.40 (16)  & \\
& $\Gamma$ ($10^{-6}$ eV)   & 1.4 (1)&  {\rm n.a.}    \\[6pt]
\hline
\\[1pt]
$^8$Li($3^+\rightarrow 2^+$) & E2        &  6.09 (10) & \\
& $\Gamma$ ($10^{-4}$ eV)    & 2.5 (1)  &   {\rm n.a.}      \\[6pt]
$^8$B($3^+\rightarrow 2^+$)  & E2        &  8.64 (23)    & \\
& $\Gamma$ ($10^{-4}$ eV)    & 5.8 (3) &  {\rm n.a.}        \\[6pt]
\hline
\\[2pt]
$^9$Li($\frac{1}{2}^-\rightarrow \frac{3}{2}^-$)& E2 & 3.69 (9)  & \\
& $\Gamma$ ($10^{-4}$ eV)                       & 7.7 (4)  &   {\rm n.a.}  \\[6pt]
\hline
\\[2pt]
$^9$Be($\frac{5}{2}^-\rightarrow \frac{3}{2}^-$)& E2  & 12.39 (15) & \\
& $\Gamma$ ($10^{-3}$ eV)                       & 1.7 (1)  &  1.89 (14)      \\[6pt]
\hline
\hline
\end{tabular}
\end{center}
\end{table}

\section {Conclusions}
\label{sec:concl}

In this work we have reported GFMC results for EM moments and
transitions of $A\leq 9$ nuclei.  The calculations of m.m.'s and $M1$
transitions account for the effect of two-body EM currents, for which
we considered two models: i) the SNPA model described in
Refs.~\cite{Mar05,MPPSW08}, and ii) the pionful $\chi$EFT EM operator
derived in~\cite{Pastore08,PGSVW09,Piarulli12}.
The goals of this work were to continue the study initiated in
Refs.~\cite{PPW07,MPPSW08}, by extending the calculations 
to systems with $A>7$, and to test the $\chi$EFT two-body
EM current operator within a hybrid context.
Both models describe the long-range behavior of the two-body EM current
in terms of OPE contributions. These pseudoscalar terms constitute
the major contribution to the total MEC correction.
The models also include, {\it albeit} in different formulations, the effects
due to currents involving $\Delta$ isobar degrees of freedom.
While the SNPA current does not involve free parameters, the $\chi$EFT
EM operator invokes a number of unknown LECs which have been fixed to
reproduce EM observables in the $A=2,3$ nuclei.
This additional freedom is probably responsible for the closer agreement
with experiment given by the $\chi$EFT formulation compared to the SNPA
model.
In particular, both the isoscalar and isovector MEC corrections for the
m.m.'s are closer to experimental data when calculated with the $\chi$EFT
EM currents.
Nevertheless, we find that the two models are in good qualitative
agreement and both support the necessity of adding MEC corrections
to reach agreement with the experimental data.

In view of the study presented in Ref.~\cite{Utsuno04} and references therein,
we have paid special attention to the $^9$Li, $^9$C mirror nuclei.
The experimental value for the isoscalar spin expectation value,
$\left< \sigma_z \right> = 1.44$, has been considered `anomalous' and
various explanations, including a broken mirror symmetry, have been suggested.
We find that calculations of $\left< \sigma_z \right>$ obtained assuming
mirror symmetry are close to the single-particle estimate of 1, even if
MEC contributions are included.
When the mirror nuclei wave functions are constructed individually,
including the appropriate Coulomb differences in both the starting VMC
trial function and GFMC propagator, and thus breaking mirror symmetry,
both the IA and MEC components of $\left< \sigma_z \right>$ are increased
in the right direction, giving a result that is consistent with the
experimental value, and indicating that both broken symmetry and MEC
contributions are required. However, the error associated with this last
calculation of the $\left< \sigma_z \right>$ value is too large to make a
definitive statement at this stage.

Finally, we have studied a number of EM transitions induced by
the $M1$ and $E2$ operators.  After including the MEC contributions,
the calculated $M1$ transition widths are in excellent agreement with
the experimental data for the $A=6,7$ nuclei and the 
$A=8$~($1^+\!\rightarrow\!2^+$) cases, while theory underpredicts the data 
in the $A=8$~($3^+\!\rightarrow\!2^+$) and $A=9$ cases.  
However, the latter $A=8$ transitions have large experimental
errors, so new precise precise measurements for these as well as for the as yet
unmeasured $^9$Li($\frac{3}{2}^-\!\rightarrow\!\frac{1}{2}^-$) transition would be most useful.
For the $E2$ transitions we provide only IA results.  While two-body corrections
to the IA charge operator have been derived in both SNPA~\cite{Riska89,Schiavilla90} and
$\chi$EFT~\cite{Pastore11}, nevertheless their leading contribution due to
OPE is expected to be small, because the associated operator vanishes in 
the static limit.  The present IA calculations appear to describe satisfactorily
the $E2$ transition widths for both the 
$^7$Li($\frac{1}{2}^-\!\rightarrow\!\frac{3}{2}^-$) and
$^9$Be($\frac{5}{2}^-\!\rightarrow\!\frac{3}{2}^-$) cases.

\acknowledgments

R.S. would like to thank the T-2 group in the Theoretical Division at LANL,
and especially J. Carlson and S. Gandolfi, for the support and warm hospitality
extended to him during a sabbatical visit in the Fall 2012, during which
part of this work was completed.  The many-body calculations were performed
on the parallel computers of the Laboratory Computing Resource Center,
Argonne National Laboratory, the computers of the Argonne Leadership Computing
Facility (ALCF) via an INCITE grant, and of the National Energy Research 
Scientific Computing Center (NERSC) at Livermore.  
This work is supported by the U.S.~Department of Energy, Office of Nuclear 
Physics, under contracts No.~DE-AC02-06CH11357 (S.P., S.C.P., and R.B.W.) 
and No.~DE-AC05-06OR23177 (R.S.) and under the NUCLEI SciDAC-3 grant.

\end{document}